\documentclass[journal,comsoc]{IEEEtran}
\usepackage{color}
\def\be{ \begin{eqnarray} }
\def\ee{ \end{eqnarray} }
\usepackage{enumitem}
\usepackage{amsthm}
\newtheorem*{remark}{Remark}
\usepackage{amsmath}

\usepackage[linesnumbered,ruled]{algorithm2e}

\usepackage[noend]{algpseudocode}
\newtheorem{problem}{Problem}
\usepackage{amsfonts}
\usepackage{subfigure}
\usepackage{cite}

\usepackage[T1]{fontenc}

\ifCLASSINFOpdf
  \usepackage[pdftex]{graphicx}
\else
  \usepackage[dvips]{graphicx}
\fi
\usepackage{amsmath}
\usepackage[cmintegrals]{newtxmath}

\usepackage[colorlinks,citecolor=magenta,urlcolor=blue,bookmarks=false,linkcolor=blue,hypertexnames=true]{hyperref}

\begin{document}
%
\title{\LARGE A Virtual Network Customization Framework for Multicast Services in NFV-enabled Core Networks}
%
%
%

\author{Omar~Alhussein,~\IEEEmembership{Student~Member,~IEEE,}
        Phu~Thinh~Do,~\IEEEmembership{Member,~IEEE,}
        Qiang~Ye,~\IEEEmembership{Member,~IEEE,}
        Junling~Li,~\IEEEmembership{Student~Member,~IEEE,}
        Weisen~Shi,~\IEEEmembership{Student~Member,~IEEE,}
        Weihua~Zhuang,~\IEEEmembership{Fellow,~IEEE,}
        ~Xuemin~(Sherman)~Shen,~\IEEEmembership{Fellow,~IEEE,}
        Xu~Li,~\IEEEmembership{Member,~IEEE,}
        and~Jaya~Rao,~\IEEEmembership{Member,~IEEE}
\thanks{Omar~Alhussein,~Junling~Li,~Weisen~Shi,~Weihua~Zhuang,~and~Xuemin~(Sherman)~Shen are with the Department of Electrical and Computer Engineering, University of Waterloo, Waterloo, ON, Canada, N2L 3G1 (emails: \{oalhusse,~j742li,~w46shi,~wzhuang,~sshen\}@uwaterloo.ca).}
\thanks{Phu Thinh Do is with Institute of Research and Development, Duy Tan University, Da Nang 550000, Vietnam (email:dopthinh@gmail.com)}
\thanks{Qiang Ye is with the Department of Electrical and Computer Engineering and Technology, Minnesota State University, Mankato, MN 56001 USA (e-mail: qiang.ye@mnsu.edu).}
\thanks{Xu Li and Jaya Rao are with Huawei Technologies Canada Inc., Ottawa, ON, Canada, K2K 3J1 (emails: \{Xu.ica,jaya.rao\}@huawei.com).}
\thanks{This paper is presented in part at IEEE Globecom 2018 \cite{Alhussein}.}
}

%
%

\markboth{Accepted to IEEE Journal on Selected Areas in Communications}%
{O. Alhussein \MakeLowercase{\textit{et al.}}: A Virtual Network Customization Framework for Multicast Services in NFV-enabled Core Networks}

%

\IEEEpubid{\copyright~2020 IEEE}


\maketitle
\begin{abstract}
The paradigm of network function virtualization (NFV) with the support of software defined networking (SDN) emerges as a promising approach for customizing network services in fifth generation (5G) networks. In this paper, a multicast service orchestration framework is presented, where joint traffic routing and virtual network function (NF) placement are studied for accommodating multicast services over an NFV-enabled physical substrate network. First, we investigate a joint routing and NF placement problem for a single multicast request accommodated over a physical substrate network, with both single-path and multipath traffic routing. The joint problem is formulated as a mixed integer linear programming (MILP) problem to minimize the function and link provisioning costs, under the physical network resource constraints, flow conservation constraints, and NF placement rules; Second, we develop an MILP formulation that jointly handles the static embedding of multiple service requests over the physical substrate network, where we determine the optimal combination of multiple services for embedding and their joint routing and placement configurations, such that the aggregate throughput of the physical substrate is maximized, while the function and link provisioning costs are minimized. Since the presented problem formulations are NP-hard, low complexity heuristic algorithms are proposed to find an efficient solution for both single-path and multipath routing scenarios. Simulation results are presented to demonstrate the effectiveness and accuracy of the proposed heuristic algorithms.
\end{abstract}


%
\IEEEpeerreviewmaketitle

\section{Introduction}
\label{sec:introduction}
%
%
%
%
\IEEEPARstart{I}{n} recent years, communication networks have experienced a radical and fundamental change in the way they are designed and managed. This shift is mainly due to two important paradigms, namely software defined networking (SDN) and network function virtualization (NFV). SDN and NFV are considered to be important technical approaches for future communication networks, representing two driving innovation platforms in the upcoming fifth generation (5G) era.
Via NFV and SDN, an approach for establishing service-customized networks has a potential to greatly resolve the complexity in conventional networks and to meet new demands and use cases \cite{Afolabi2018,Zhang2017b,Huang2015,Shan2018,Peng2018,Shi2018}. A service-customized network provides traditional connectivity between a set of terminals, and allows for the deployment of abstract network applications in the data plane. 
An NFV-enabled network service can be represented by a logical topology, called a \emph{network function (NF) chain}, which specifies a set of virtual NFs that are orchestrated and deployed along the chosen routes from the source(s) to the destination(s), with some properties and dependencies satisfied. Some examples of virtual NFs include cache, transcoder, firewall, 5G evolved packet core, wireless access network (WAN) optimizer and network address translator (NAT), which can be hosted and dynamically employed at commodity servers and data center (DC) nodes (called \textit{NFV nodes}).
In this approach, the service provider (SP) requests a network service to satisfy certain expected demands and requirements as per the service-level agreement (SLA). In turn, the responsibility of the service orchestrator or the network operator (NO) is to efficiently design such a virtual network, and place it on the physical substrate network. The NO aims at reducing the provisioning cost, and maximizing the utilization of its own networks, while simultaneously meeting the binding SLA.
\IEEEpubidadjcol

\begin{figure}
\centering
\includegraphics[width=0.5\textwidth]{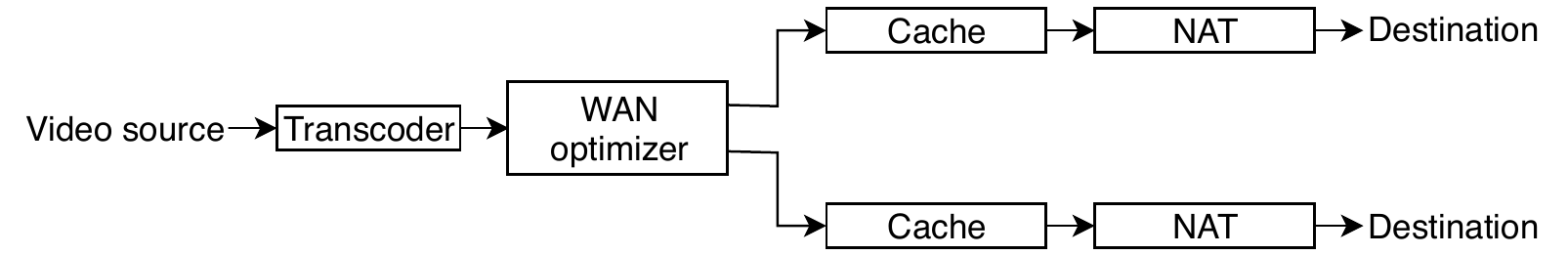}
\caption{Illustraion of a multicast NF chain for a basic video streaming service.}
\label{fig:intro:motivationNFV}
\end{figure}

There is a considerably growing demand for services of \textit{multicast nature} such as video streaming, multi-player augmented reality games, and file distribution. For instance, it is estimated that 82 percent of the consumers Internet traffic will be attributed to video streaming \cite{8858585}. In addition, the multicast mode of communication can reduce the bandwidth consumption in backbone networks by over 50\% in contrast to the unicast mode \cite{Malli1998}. In an NFV-enabled network, a multicast service can deploy several NF instances in the data plane, whereas a routing policy forwards the traffic from the source to each destination while traversing the NFs instances as per the NF chain. Fig. \ref{fig:intro:motivationNFV} illustrates an example of a multicast NF chain for a typical video streaming service that distributes content to several destinations. If a multicast service requires connecting some terminal nodes without intermediate NFs, the optimal routing that reduces the link provisioning cost can be found by constructing a Steiner tree or one of its variants. A Steiner tree is a generalization of the minimum spanning tree (MST) which finds the subset of weighted edges (and nodes) that connect all vertices in a graph with the minimum possible link provisioning cost. Constructing a Steiner tree is an NP-hard problem \cite{Vazirani2003,Gilbert1968}. However, polynomial time approximation algorithms exist to construct a Steiner tree.
With the emergence of SDN providing a global view of the physical network topology and network states, Steiner tree-based routing approaches become feasible \cite{Zhao2014,Qiao2015,Blendin2015}. However, such approaches do not incorporate NFs in their formulation, and cannot be extended directly to the joint multicast routing and NF placement problem. Practically, there exist a massive number of NF placement configurations, each of which requires a multicast routing topology construction (e.g., one instance of such configurations is shown in Fig. \ref{fig:intro:motivationNFV}). The NF placement and multicast routing are correlated, which leads to technical challenges for orchestrating a single NFV-enabled multicast service. Selecting just enough NFV nodes for NF placement inevitably increases the link provisioning cost for building an appropriate multicast routing topology; Conversely, instantiating NF instances on more NFV nodes may yield a decreased link provisioning cost with traffic load balancing at the expense of an increased function provisioning cost. Therefore, how to balance the tradeoff between link and function provisioning costs is a challenging issue.
Our objective is to determine an optimal NF placement with multicast traffic routing to minimize the overall function and link provisioning costs. Another major challenge stems from the fact that multiple network services share the network substrate. As the network substrate has limited transmission and processing resources, the efficiencies of embedding multiple service requests interplay. Prioritizing a low-rate network service for embedding can fragment the network resources, thereby hindering other high-rate network services from being successfully (or efficiently) embedded.
Some recent studies address the orchestration of NFV-enabled multicast service to minimize the function and link provisioning costs \cite{Zhang2015,Zhang2016,Zeng2016,Xu2017a,Kuo2017,Alhussein}. However, most existing works assume a design scenario where all NFs are hosted in one NFV node for each network service and multicast replication points can occur only after the deployment of NFs. More realistic and flexible design (e.g., one NFV node is only capable of hosting specific types of NFs, multipath routing is enabled between NFs) can make the existing solutions not feasible or not optimized. More details of relevant literature are discussed in Section \ref{sec:related_work}.

In this paper, we present an optimization framework for the orchestration of multiple multicast service requests over the physical substrate. First, we study a joint multicast traffic routing and NF placement problem for a single service request to minimize the function and link provisioning costs, under the physical processing and transmission resource constraints, flow conservation constraints, and NF placement rules. For practical applications, our formulated problem focuses on \textit{flexible} multicast routing and NF placement (embedding), where we allow \textit{one-to-many} and \textit{many-to-one} NF mappings. That is, several NF instances can be hosted at one NFV node if permissible, and one type of NF can be replicated and deployed on different NFV nodes as NF \textit{instances} to serve different sets of destinations, 
thereby providing considerable flexibility in the deployment of network services. Furthermore, our formulated problem incorporates both single path and multipath traffic routing between the embedded NF instances. Since the formulated problem is NP-hard, we devise a low-complexity heuristic algorithm to obtain an efficient and flexible solution, based on a key-node preferred minimum spanning tree (MST) approach;
Second, we consider a general scenario of placing multiple service requests over the physical network. Since multiple services compete with each other to be hosted on a substrate network with limited resources, we accept the services which maximize the aggregate throughput with least provisioning cost. We formulate an MILP that jointly determines multicast topologies for multiple service requests, where we find the combination of network services that maximize the aggregate network throughput while minimizing the overall function and link provisioning costs. Moreover, we develop a simple heuristic algorithm that prioritizes the service requests, aiming at maximizing the aggregate throughput with the minimum overall provisioning cost.

The rest of the paper is organized as follows. Section \ref{sec:related_work} gives an overview of related work. Section \ref{sec:System Model} presents the system model under consideration, which includes the representation of the physical network, NFs, and multicast service requests.
Section \ref{sec:problem_definition} addresses the joint NF placement and routing problems for multicast services with multipath routing for both single-service and multi-service cases.
Section \ref{sec:problem_formulation} presents MILP formulations for a single-service multipath scenario, and a generalized multi-service multi-path scenario, respectively. Section \ref{sec:heuristic} proposes simple modular heuristic algorithms to address the complexity of the resultant MILP formulations. Simulation results are presented in Section \ref{sec:simulations}, and concluding remarks are drawn in Section \ref{sec:conclusions}. A list of important notations is given in Table \ref{table:notations}.

\section{Related Work}
\label{sec:related_work}
SDN provides a global network view and centralized control over the substrate network topology with the associated network states. Some existing works focus on constructing efficient centralized routing topologies for multicast services without NFs \cite{Huang2014,Zhao2014,Blendin2015,Qiao2015}, which mainly rely on constructing a Steiner tree (or one of its variants) to reduce the transmission resource consumption, network state consumption, end-to-end (E2E) delay, or to improve the reliability. For instance, Zhao \emph{et al.} propose an SDN-based video conferencing solution by constructing a minimum-delay Steiner tree with iterative enhancements \cite{Zhao2014}. 
To deal with the increased network state consumption due to multicasting, a Steiner tree  that jointly minimizes the number of links and branch nodes (with network states) is established, assuming unicasting across links with no branches~\cite{Huang2014}.
The state-of-the-art SDN-enabled multicast routing approaches cannot be adopted or modified directly to incorporate the NF placement.

The virtual NF placement and traffic routing problem has been extensively studied for the unicast service case \cite{Mehraghdam2014,Bari2015,Luizelli2015,Ghaznavi2017,Xie2016,Zhang2016_sector,Zhang2017b,Li,Ye2018}. For the multicast service case, a simple approach is to apply the unicast-based NF placement and routing approaches to each source-destination pair independently, which leads to a waste of network resources with a large service provisioning cost. There are relatively few works in the joint multicast routing and NF placement for multicast services \cite{Zhang2015,Zhang2016,Zeng2016,Xu2017a,Kuo2017,Alhussein}, where one needs to jointly build a multicast topology and place the NFs.
Zhang \emph{et al.} consider the joint routing and placement for NFV-enabled multicast requests, under the assumption that all NFs should be served by only one NFV node \cite{Zhang2015}, which is extended for a multiple NFV node case under the assumption of an uncapacitated substrate network \cite{Zhang2016}. It is assumed that multicast replication points occur only after deployment of NFs in the constructed multicast topology, which reduces the degree of flexibility of the orchestration framework.
Zeng \emph{et al.} jointly consider placement, multicast routing, and spectrum assignment for a fibre optical network, where the objective is to minimize the function, link, and frequency spectrum provisioning costs  \cite{Zeng2016}.
%
The heuristic solution clusters each group of destinations that share one specific NF. Then, the solution is divided into two steps to allocate the NF in an NFV node for each cluster, and to find the MST that spans the source, the allocated NF, and the destinations. Similarly, it is assumed that, for each multicast service, the traffic flowing to each destination is processed by one NF.

Xu \emph{et al.} consider that multiple NFV nodes can host all types of NFs \cite{Xu2017a}. For each source-destination pair, the multicast stream needs to pass through only one NFV node where all NFs are placed before arriving at each destination. Since the destinations are distributed in a large area, the algorithm allows activating multiple NFV nodes, where each NFV node is responsible for a subset of destinations. However, there is possibility that some NFV nodes can only host specific types of NFs due to hardware-based or subscription-based restrictions \cite{Zeng2016,Bari2015}.
A recent work tackles the so-called service forest problem for the traffic routing and NF placement of multiple service requests \cite{Kuo2017}. For a generalized scenario where each source-destination pair requires multiple service requests, the last NF is placed on the physical substrate first, which divides the multicast topology into two parts. The first part is from the source to the last NF, while the second part from the last NF to all destinations. The first part is solved by the so-called $k$-stroll algorithm. In the second part, a minimum Steiner tree approximation connects the last NF to all destinations. It is assumed that multicast replication points at the topology occur always after the last NF, and an exhaustive search finds the best place to host the last NF among all candidate NFV nodes.

To address some research gaps, we aim at developing a \textit{flexible} multicast routing and NF placement framework for both single-service and multi-service scenarios. We consider that multicast replication points can occur before and after the deployment of NF instances, thereby providing flexibility for topology customization which is particularly crucial for geographically distributed NF chains (such as in NFV-enabled core networks). In addition, multipath routing between the embedded NFs is incorporated to improve the utilization of the network substrate's resources.


\section{System Model}
\label{sec:System Model}
\subsection{Physical Substrate Network}
Consider a physical substrate network, $\mathcal{G}=(\mathcal{N},\mathcal{L})$, where $\mathcal{N}$ and $\mathcal{L}$ are the set of nodes and links. The nodes can be switches (represented by set $\mathcal{F}$), commodity servers and data center (DC) nodes (namely NFV nodes, represented by set $\mathcal{M}$), \textit{i.e.}, $\mathcal{N}=\mathcal{F}\cup\mathcal{M}$. Switches are capable of forwarding and replicating traffic, and NFV nodes are capable of hosting and operating NFs. We assume that each NFV node has a forwarding capability, and has available processing rate $C(n)$, $n \in \mathcal{M}$, in packet per second (packet/s) \cite{6550275,Ghodsi:2012:MFQ,Ghodsi:2011:DRF}. Moreover, an NFV node is capable of provisioning a number of NFs simultaneously as long as the available processing rate satisfies the NF processing requirements. Each physical link has a limited transmission rate $B(l)$, $l \in \mathcal{L}$, in packet/s. 



\subsection{NFs}
We represent all the NF types by set $\mathcal{P}$, where a specific type, $p\in \mathcal{P}$, resembles some virtual functionality (e.g., IDS, compression, proxy, and LTE packet gateway).
We further associate NFV node $n\,(\in \mathcal{M})$ with a set of admittable NF types using an indicator function $k_{ni}\in\{0,1\}$, where $k_{ni}=1$ if NFV node $n\,(\in \mathcal{M})$ can admit function $f_{i}\,(\in \mathcal{P})$.

\subsection{Multicast NF Chains}
Let the set of all multicast service requests be denoted by $\mathcal{R}$. A multicast service, $r \in \mathcal{R}$, is described by a multicast NF chain, represented by a weighted acyclic directed graph, 
\begin{equation}
 S^{r}=(s^{r},\mathcal{D}^{r},\mathcal{V}^{r},\overline{d^r}),~~ r \in \mathcal{R}
\end{equation}
where $s^{r}$ and $\mathcal{D}^{r}$ represent the source node and the set of destinations, $\mathcal{V}^{r} = \{f^{r}_{1},f^{r}_{2},\dots f^{r}_{|\mathcal{V}|}\}$ represents the set of functions that have to be traversed in an ascending order for every source-destination pair, and $\overline{d^r}$ is the data rate requirement in packet/s.
Each NF $f^{r}_{i}$ requires a processing rate $C(f^{r}_{i})$, $i\in\{1,\dots,|\mathcal{V}^{r}|\}$, in packet/s. An NF instance belongs to one service request, and it cannot be shared among multiple NF chains \cite{Kuo2017,Zhang2016,Zhang2015,Mehraghdam2014}.

\section{Problem Definition}
\label{sec:problem_definition}
We investigate the orchestration of multiple multicast services over the physical substrate network in two sequential problems. In the first problem, joint multipath-enabled multicast routing and NF placement are studied for a single service. 
Given the substrate network and the service description of a multicast NF chain, we intend to design an embedded multicast topology for the NF chain. For each source-destination pair in the embedded topology, all NF types should be traversed in a specified order. Therefore, the first problem consists of two joint subproblems: (i) NF placement on the physical substrate, and (ii) multicast traffic routing design from the source to the destinations, passing through a sequence of the required NFs. Note that multipath routing is enabled for the paths between embedded NFs to increase the flexibility of topology customization and improve the physical resource utilization especially for geographically-dispersed large-scale core networks.
Our objective is to minimize the function and link provisioning costs in determining an optimal embedded multicast topology. However, the minimization of both costs are conflicting. Instantiating a large number of NF instances at more network locations achieves a balanced traffic load at the expense of an increased function provisioning cost, whereas fewer NF instantiations reduce the function provisioning cost at the expense of less load balancing and inefficient network resource exploitation.
\begin{figure}
\centering
\includegraphics[width=0.5\textwidth]{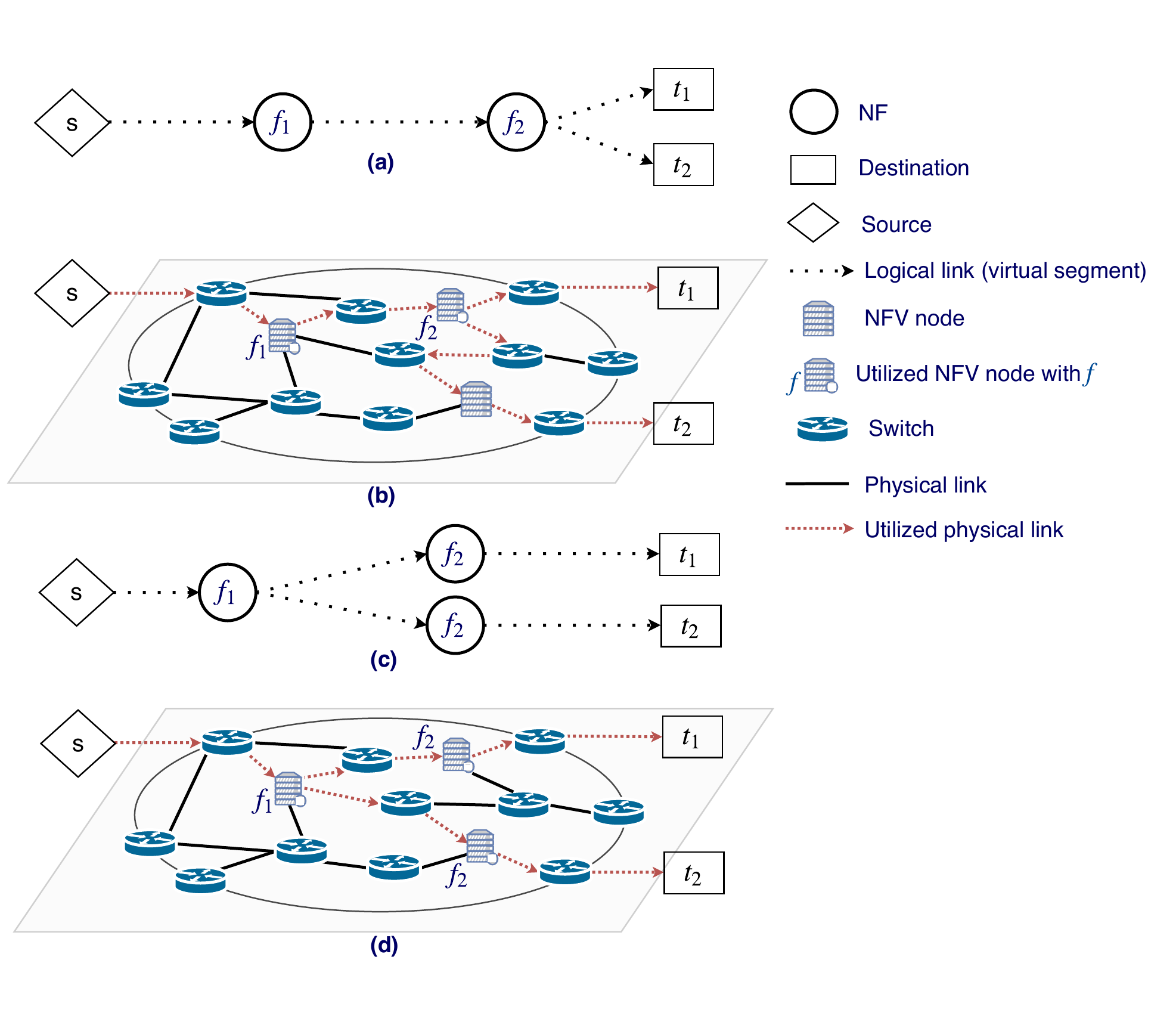}
\caption{Comparison of flexible and non-flexible embedding for a multicast request. (a) Topology of NF chain request; (b) Embedding result of NF chain on network substrate with 11 links due to the non-flexible scheme; (c) Modified topology of NF chain request due to the flexible scheme; (d) Embedding result of modified NF chain on network substrate with 10 links due to the flexible scheme.}
\label{fig:scenario_problem1}

\end{figure}
Therefore, it is required to balance the tradeoff to minimize the overall provisioning cost for the NF chain embedding. The first problem is defined as follows:

\begin{problem}
To determine an optimal multipath-enabled multicast topology for an NF chain to minimize the function and link provisioning costs, with all the required NFs traversed in order and with the required processing and transmission resource constraints satisfied. 
\end{problem}

In the second problem, we study a joint multicast routing and NF placement problem for a multi-service scenario. 
NFV allows multiple NF chains to run over a common network substrate. However, as the network resources are limited, multiple NF chains may not be accepted on the substrate network simultaneously. We need to decide which service requests should be embedded such that the aggregate throughput is maximized, while the function and link provisioning costs are minimized. We consider the static NF chain embedding for the multi-service scenario, where all service requests are available a priori. The online NF placement and routing in which different types of services arrive and being embedded dynamically is subject to our future research. Therefore, the second problem is defined as follows:

\begin{problem}
To find an optimal combination of multicast NF chains that maximizes the aggregate throughput of the network substrate, while minimizing the respective function and link provisioning costs.
\end{problem}

We consider that NFs and virtual links have one-to-many and many-to-one mapping with physical NFV nodes and links, respectively, whereby each NF instance can serve a subset of destinations, and the deployment of NF instances can occur after packet replication in the multicast topology. Such practical consideration achieves higher flexibility and efficiency in the routing and placement processes for largely distributed networks, since destinations may be geographically far away. Restricting all destinations to share the same set of NF instances can be inefficient for transmission resource limited scenarios. When the destinations are geographically dispersed, an efficient solution is to duplicate and deploy the function instances close to each of the destinations for a reduced link provisioning cost due to flexible routing. We give an illustration in Fig. \ref{fig:scenario_problem1}, where the logical topology of an NF chain request with two NFs and two destinations (i.e., $t_1,t_2\in \mathcal{D}$) are embedded onto the substrate network. Assume that each physical link can be used only once. With a non-flexible scheme, NFs cannot be replicated and multicast replication points can exclusively occur after the deployment of the last NF. Hence, the embedding result of the multicast request (Fig. \ref{fig:scenario_problem1}(a)) on the physical substrate requires 11 links as shown in Fig. \ref{fig:scenario_problem1}(b). With a flexible scheme, the NF chain topology is modified, as shown in Fig. \ref{fig:scenario_problem1}(c), where the respective embedding result on the physical substrate requires 10 links as shown in Fig. \ref{fig:scenario_problem1}(d).

%
%

\section{Problem Formulation}
\label{sec:problem_formulation}
We first present the problem formulation for a single-service multipath scenario, followed by a generalized multi-service multipath scenario.

\subsection{Single-Service Multipath Scenario}
\label{sec:problem_formulation:multipath_single}

\begin{figure}[htbp]
\centering
\includegraphics[width=0.5\textwidth]{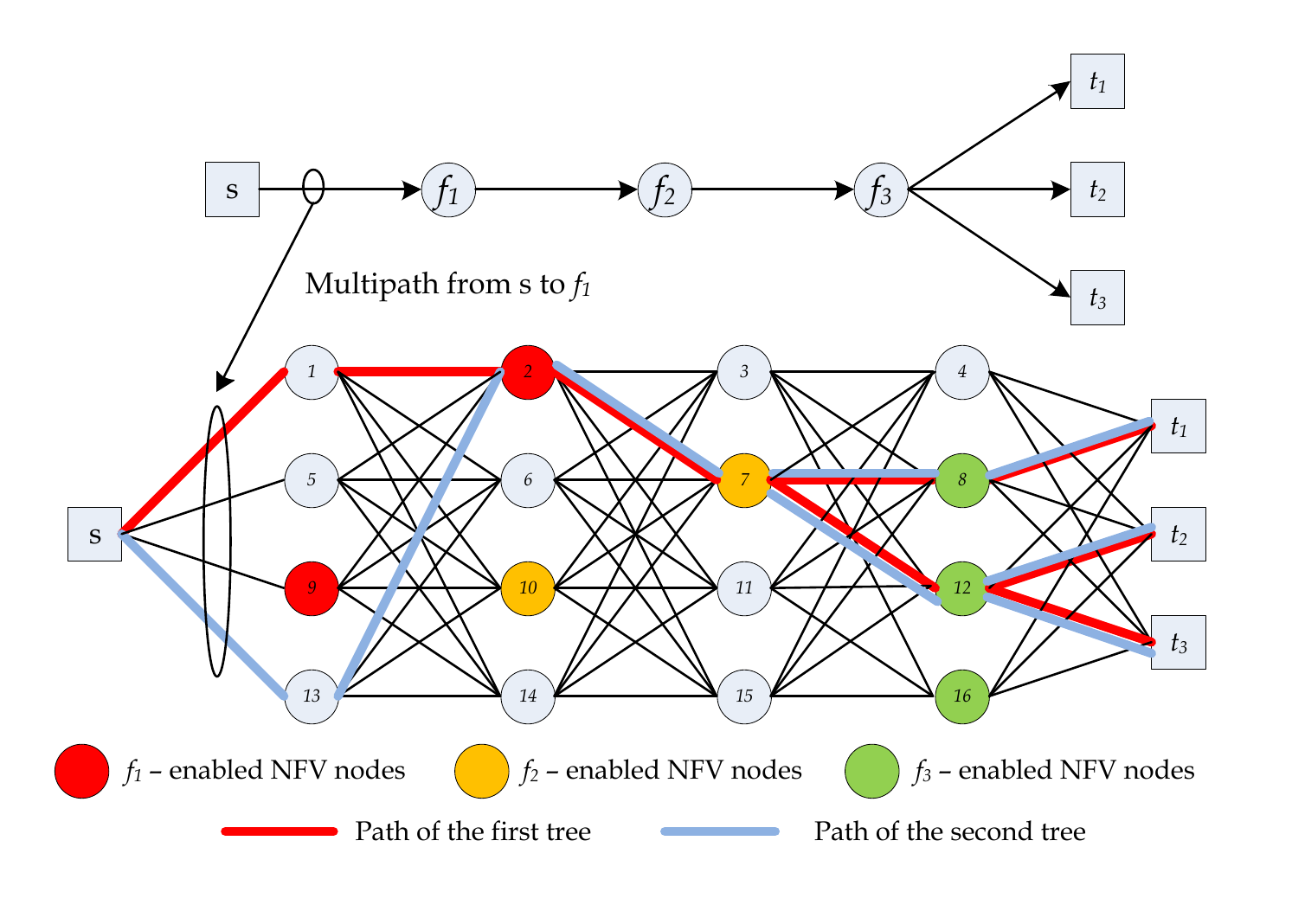}
\caption{Two Steiner trees that share the same source, traversed functions, and destinations.}
\label{fig:multipath_example}
\end{figure}

\begin{table*}
\centering
\caption{\label{table:notations}
List of important symbols for the optimal single- and multi-service formulations.}
 \begin{tabular}{||c | l ||} 
                  \multicolumn{2}{c}{\textbf{Network Substrate}} \\
 \hline \hline
 $\mathcal{G}(\mathcal{N},\mathcal{L})$ & Network substrate $\mathcal{G}$, where $\mathcal{N}$ and $\mathcal{L}$ are the set of nodes and links \\ 
 \hline
 $\mathcal{F}$ & The set of switches in $\mathcal{N}$ $(\mathcal{F}\subseteq\mathcal{N})$  \\
 \hline
 $\mathcal{M}$ & The set of NFV nodes (commodity servers) in $\mathcal{N}$ $(\mathcal{M}\subseteq\mathcal{N})$ \\
 \hline
 $C(n)$ & The available processing rate for node $n$ ($\in \mathcal{M}$) in packet/s  \\
  \hline
 $B(l) $ & The available transmission rate for link $l$ ($\in\mathcal{L}$) in packet/s \\ 
 \hline
                   \multicolumn{2}{c}{\textbf{NF and Multicast Service Requests}} \\
\hline
$\mathcal{R}$ & The set of multicast service requests \\
\hline
$s^{r}$ & The source node of service request $r$ ($\in \mathcal{R}$)  \\
\hline
$\mathcal{D}^{r}$ & The set of destinations of service request $r$ \\
\hline
$f^{r}_{i}$ & The NF with index $i$ for service request $r$ ($\in \mathcal{R}$) \\
\hline
$\mathcal{V}^{r}$ & The set of all NFs in service request $r$ ($\in \mathcal{R}$), $\mathcal{V}^{r}=\{f^{r}_{1},f^{r}_{2},\dots f^{r}_{|\mathcal{V}^{r}|}\}$ \\
\hline
$S^{r}$ & The representation of the service request $r$ ($\in \mathcal{R}$), $S^{r}=(s^{r},\mathcal{D}^{r},\mathcal{V}^{r},\overline{d^r})$ \\
\hline
$C(f^{r}_{i})$ & The required processing rate of NF $f_{i}^{r}$ in packet/s \\
\hline
$\overline{d^r}$ & The data rate requirement for service request $r$ ($\in \mathcal{R}$) in packet/s \\
\hline
$k_{ni}$ & The function to indicate whether NF $f_i$ is admittable at node $n$\\
\hline
                  \multicolumn{2}{c}{\textbf{Single-service and Multi-service Optimal Formulations}} \\
\hline
$\mathcal{S}_{m}^{n}$  &  The set of integers from $m$ to $n$ ($\mathcal{S}_{m}^{n}=\{m, m+1, \ldots, n\}$ with $m,n \in \mathbf{Z_+}$) \\
\hline
$J^{r}$ & The number of trees (maximum number of multipath routes) for service request $r$ \\
\hline
$d^{jr}$ & Variable: the data rate emanating from the tree $j$ of service request $r$\\
\hline
$z_{ni}^{r}$ & Variable: indicates that an instance of $f_i^{r}$ is deployed on node $n$ for service $r$\\
\hline
$u_{nit}^{r}$ & Variable: indicates an instance of $f_i^{r}$ is hosted at node $n$ for traffic flowing towards destination $t$ ($\in \mathcal{D}^{r}$) \\
\hline
$x_{li}^{jr}$  & Variable: indicates that link $l$ is used for forwarding traffic for service $r$ from $f_i^{r}$ to $f_{i+1}^{r}$ in tree $j$\\
\hline
$y_{lit}^{jr}$ & Variable: indicates that link $l$ is used for forwarding traffic for service $r$ from $f_i^{r}$ to $f_{i+1}^{r}$ in tree $j$ for destination $t$ \\
\hline
$\pi^{jr}$ & Variable: indicates that tree $j$ of service request $r$ is activated \\
\hline
$d^{jr}$ & Variable: the fractional transmission rate of tree $j$ of service request $r$\\
\hline
$\gamma_{li}^{jr}$ & Variable: the rate in tree $j$ over link $l$ to deliver traffic from $f_i^{r}$ to $f_{i+1}^{r}$ for service $r$\\
\hline
$\alpha,\,\beta$ & The weighting coefficients to reflect the importance level of the two objectives in \eqref{eq14}, with $\alpha+\beta=1,\,\alpha,\beta \geq 0$  \\
\hline

    \multicolumn{2}{c}{\textbf{Specific to Multi-service Optimal Formulation}} \\
\hline
$\mathbb{R}$ & The achieved aggregate throughput from a set of services $\mathcal{R}$ \\
\hline
$R^{r}$ & The achieved throughput from admitting service request $r$ \\
 \hline
$\rho^{r}$ & Variable: Indicates whether service request $r$ is admitted to the network substrate \\
\hline
\hline
\end{tabular}
\end{table*}


To establish a joint multipath-enabled multicast routing and NF placement framework for a single-service, let $J^r$ denote the maximum number of multicast trees to deliver multicast service $r$ ($\in \mathcal{R}$) from the source to the destinations\footnote{Note that the superscript $r$ which is used in the single-service formulation (subsection \ref{sec:problem_formulation:multipath_single}) can be dropped since $|\mathcal{R}|=1$. However, it is necessary as we develop the MILP formulation for the multi-service scenario in subsection \ref{sec:problem_formulation:multiservice}.}. Each tree emanates from the source and passes by the same set of traversed NFV nodes and destinations. Fig. \ref{fig:multipath_example} illustrates one multicast request with three destinations and three functions. We use two trees (i.e., $J^{r}=2$) to support up to two multipath routing paths. If $f_1$ is embedded on node $2$ for both trees, we have multipath routes from source node $s$ to node $2$, and the two trees converge afterwards. When $J^{r} = 1$, the problem reduces to the single-path routing case. The maximum number of multicast trees  (resembling the maximum possible number of multipath routes) $J^{r}$ is an input to the problem formulation. The formulation allows the multicast trees to overlap with each other, and can be deactivated if needed. In what follows, we describe the details of the MILP formulation for the single-service scenario.
%
%
%

Let $\mathcal{S}_{m}^{n}$ denote the set of integers from $m$ to $n$ ($>m$), i.e., $\mathcal{S}_{m}^{n} \triangleq \{m, m+1, \ldots, n\}$ with $m,n \in \mathbf{Z_+}$. Define binary variable $x_{li}^{jr} \in \{0,1\}$, where $x_{li}^{jr} = 1$ indicates that link $l$ $(\in \mathcal{L})$ is used for forwarding traffic for service $r$ in multicast tree $j$ from $f_i^{r}$ to $f_{i+1}^{r}$ where $i \in \mathcal{S}_1^{|\mathcal{V}^r|-1}$; Binary variable $x_{l0}^{jr} = 1$ indicates that link $l$ carries traffic from $s^r$ to $f_1^{r}$, and $x_{l|\mathcal{V}|}^{jr} = 1$ indicates that link $l$ carries traffic from $f_{|\mathcal{V}|}^{r}$ to any destination node $t\,(\in \mathcal{D}^r)$; Define $y_{lit}^{jr} \in \{0,1\}$, where $y_{lit}^{jr}=1$ indicates that link $l$ is used to direct traffic for service $r$ in multicast tree $j$ from $f_i^{r}$ to $f_{i+1}^{r}$ for destination $t\,(\in \mathcal{D}^r)$; Binary variable $y_{l0t}^{jr}=1$ indicates that link $l$ is used to direct traffic for service $r$ in tree $j$ from $s^r$ to $f_1^{r}$ for destination $t$, and $y_{l|\mathcal{V}|t}^{jr}=1$ indicates that link $l$ directs traffic for service $r$ in tree $j$ from $f_{|\mathcal{V}|}^{r}$ to destination $t$.

With the definitions of $\boldsymbol{x} = \{x_{li}^{jr}\}$ and $\boldsymbol{y} = \{y_{lit}^{jr}\}$, we have
\begin{equation}
y_{lit}^{jr} \leq x_{li}^{jr}, ~~l \in \mathcal{L}, i \in \mathcal{S}_0^{|\mathcal{V}^r|}, j \in \mathcal{S}_{1}^{J^r}, t\in \mathcal{D}^r, ~ r \in \mathcal{R}.
\label{eq01}
\end{equation}
Furthermore, define binary variables $z_{ni}^{r} \in\{0,1\}$, where $z_{ni}^{r}=1$ indicates that an instance of $f_{i}^{r}$ is deployed on NFV node $n$ for service $r$ where $i \in \mathcal{S}_{1}^{|\mathcal{V}|}$, and $u_{nit}^{r} \in\{0,1\}$, where $u_{nit}^{r}=1$ indicates that an instance of $f_{i}^{r}$ is deployed on $n$ for service $r$ for destination $t$. Similarly, we have a relationship constraint between $\boldsymbol{z} = \{z_{ni}^{r}\}$ and $\boldsymbol{u} = \{u_{nit}^{r}\}$ as
\begin{equation}
    u_{nit}^{r} \leq z_{ni}^{r}, ~~ n \in \mathcal{N}, i \in \mathcal{S}_1^{|\mathcal{V}^{r}|},~t\in \mathcal{D}^{r},~ r \in \mathcal{R}.
\label{eq03}
\end{equation}

For each service $r$ ($\in \mathcal{R}$), we build $J^{r}$ multicast trees to exploit the multipath property, where each tree can provide a fractional transmission rate $d^{jr}$ of the overall required transmission rate $\overline{d^{r}}$. Therefore, to meet the total required transmission rate $\overline{d^{r}}$, we impose the constraint
\begin{equation}
\sum_{j=1}^{J^{r}} d^{jr} \geq \overline{d^{r}}, ~ r \in \mathcal{R}.
\label{eq04}
\end{equation}
Next, we incorporate the routing and placement constraint in (\ref{eq07}) to ensure that traffic flows pass from the source to multiple destinations through the NF chain. Let $f_{0}^{r}$ and $f_{|\mathcal{V}|+1}^{r}$ denote dummy functions (without processing requirements) that are placed on the source node $s^r$ and each destination node $t$ ($\in\mathcal{D}^r$), respectively. In our model, some of the multicast trees can be deactivated if needed. Consequently, we have
\begin{eqnarray}
    \nonumber & \hspace{-1cm}\sum\nolimits_{(n,m) \in \mathcal{L} } y_{(n,m)it}^{jr} - & \sum\nolimits_{(m,n) \in \mathcal{L}} y_{(m,n)it}^{jr} = \\
    & &  \begin{cases}
    u_{nit}^{r} - u_{n(i+1)t}^{r}, &\text{tree $j$ is activated}
    \\
    0, &\text{otherwise}
    \end{cases}
    \label{eq07}
\end{eqnarray}
for $n \in \mathcal{N}, i \in \mathcal{S}_0^{|\mathcal{V}^{r}|}, t \in \mathcal{D}^{r}, r \in \mathcal{R}$, where 
\begin{equation}\nonumber
u_{s0t}^{r}=1,\,u_{n0t}^{r}=0,\,\forall t\in\mathcal{D}^{r},\,n\in\mathcal{N}\backslash\{s^{r}\}
\end{equation}
\begin{equation}\nonumber
u_{t(|\mathcal{V}^{r}|+1)t}   =1,\,u_{n(|\mathcal{V}^{r}|+1)t}=0,\,\forall t\in\mathcal{D}^{r},\,n\in\mathcal{N}\backslash\mathcal{D}^{r}
\end{equation}
\begin{equation}\nonumber
u_{sit}^{r}=0,\,u_{tit}^{r}=0,\,\forall i\in\mathcal{S}_{1}^{|\mathcal{V}^{r}|},\,t\in\mathcal{D}^{r}.
\end{equation}
Define binary variable  $\pi^{jr} \in \{0,1\}$, where $\pi^{jr}=1$ indicates that tree $j$ of service $r$ is activated. Consequently, all variables related to deactivated trees (i.e., $x_{li}^{jr}$, $y_{lit}^{jr}$,  and $d^{jr}$) should be forced to zero. Therefore, we impose the constraint
\begin{align}
    \nonumber x_{li}^{jr} \leq \pi^{jr}, & ~ y_{lit}^{jr} \leq \pi^{jr}, d^{jr} \leq \pi^{jr} \overline{d^{r}}, \\
    & ~~l \in \mathcal{L}, i \in \mathcal{S}_0^{|\mathcal{V}^r|}, j \in \mathcal{S}_{1}^{J^r}, t\in \mathcal{D}^r, ~ r \in \mathcal{R}.
    \label{eq10}
\end{align}
To guarantee that the routing and placement constraint is considered only when tree $j$ of service $r$ ($ \in \mathcal{R}$) is activated, we modify \eqref{eq07} to

\begin{align}
\sum_{(n,m) \in \mathcal{L} } y_{(n,m)it}^{jr} -  & \sum_{(m,n) \in \mathcal{L}} y_{(m,n)it}^{jr} = \pi^{jr} \left(u_{n(i+1)t}^{r} - u_{nit}^{r} \right),
\nonumber\\
&
~n \in \mathcal{N}, i \in \mathcal{S}_0^{|\mathcal{V}^{r}|}, t \in \mathcal{D}^{r}, r \in \mathcal{R}.
\label{eq11}
\end{align}
Since we have $y_{lit}^{jr} \leq x_{li}^{jr}$ in \eqref{eq01}, the constraint $y_{lit}^{jr} \leq \pi^{jr}$ in \eqref{eq10} can be removed.  Thus, we re-write \eqref{eq10} as
\begin{equation}
    x_{li}^{jr} \leq \pi^{jr},~d^{jr} \leq \pi^{jr} \overline{d^r}, ~ l \in \mathcal{L}, i \in \mathcal{S}_0^{|\mathcal{V}^{r}|}, j \in \mathcal{S}_1^{J^{r}}.
    \label{eq12}
\end{equation}
Constraint \eqref{eq12} means that we consider $x_{li}^{jr}$ and $d^{jr}$ when tree $j$ is activated; otherwise, we simply set these variables to zero.
We require that exactly one instance of function $f_{i}^{r}$ is traversed for every source-destination pair, which can be expressed as
\begin{equation}
    \sum\nolimits_{n \in \mathcal{M}} u_{nit}^{r} = 1, ~ i \in \mathcal{S}_1^{|\mathcal{V}^{r}|}, t \in \mathcal{D}^{r}, r \in \mathcal{R}.
    \label{eq08}
 \end{equation}
Moreover, function $f_i^{r}$ is hosted at node $n$ only when admittable and when the resources at node $n$ are sufficient. We have
\begin{subequations}
\begin{eqnarray}
    & &\hspace{-1.3cm} \sum_{r \in \mathcal{R}} \sum_{i =1}^{|\mathcal{V}^{r}|} z_{ni}^{r} C(f_{i}^{r})  \leq C(n), ~ n \in \mathcal{M},~i \in \mathcal{S}_{1}^{|\mathcal{V}^{r}|} \label{eq:capacity_p1} \\
    & &\hspace{-1.3cm} z_{ni}^{r} k_{ni} =1, ~ n \in \mathcal{M}, i \in \mathcal{S}_{1}^{|\mathcal{V}^{r}|}, r \in \mathcal{R}
    \label{eq:subscription_p1}
\end{eqnarray}
\label{eq:subscription_capacity}%
\end{subequations}
where $k_{ni}=1$ indicates that node $n$ can admit function $f_{i}$; otherwise, $k_{ni}=0$.
%

\emph{Objectives} --- Following the relevant research on NFV-enabled service provisioning \cite{Kuo2017,Xu2017a,Zeng2016,Luizelli2015,Bari2015,Mehraghdam2014}, we aim to minimize the function (processing) and link (transmission) provisioning costs over all $J^{r}$ multicast trees for service $r$, in addition to balancing the substrate network resources in the long run as
\begin{eqnarray}
& &\hspace{-1.3cm}
    \underset{}{\min}~ \alpha \sum_{l \in \mathcal{L}}\sum_{j=1}^{J^{r}}\sum_{i =0}^{|\mathcal{V}^{r}|}\left(\frac{ d^{jr}} {B(l)} + 1 \right)x_{li}^{jr} 
    + \beta \sum_{i =1}^{|\mathcal{V}^{r}|}\sum_{n \in \mathcal{M}}\frac{C(f_i^{r})}{C(n)}z_{ni}^{r}.
    \label{eq14}
\end{eqnarray}
In (\ref{eq14}), we minimize the weighted sum of the cost of forwarding the traffic over the utilized physical links of the substrate network for all activated trees, and the cost of provisioning the NF instances in the NFV nodes. Parameters $\alpha$ and $\beta$ are the weighting coefficients to reflect the importance level of minimizing the cost of traffic forwarding and minimizing the cost of NF provisioning respectively, where $\alpha+\beta = 1$ and $\alpha,\,\beta > 0$. The terms $d^{jr}  x_{li}^{jr}/ B(l)$ and $C(f_i^{r}) z_{ni}^{r}/C(n)$ ensure load balancing over the physical links and NFV nodes \cite{Chowdhury2012}. Moreover, the term `$+1$' in \eqref{eq14} minimizes the number of links in building the multicast topology.
Denote the product term $d^{jr} x_{li}^{jr}$ in the objective function \eqref{eq14} by $\gamma_{li}^{jr}$ as
\begin{equation}
    \gamma_{li}^{jr} = x_{li}^{jr} d^{jr}, ~ l \in \mathcal{L}, i \in \mathcal{S}_0^{|\mathcal{V}^{r}|}, j \in \mathcal{S}_1^{J^{r}}, r\in\mathcal{R}.
    \label{eq06}
\end{equation}
The term, $\gamma_{li}^{jr}$, can be interpreted as the transmission rate over link $l$ to deliver traffic from $f_i^{r}$ to $f_{i+1}^{r}$ in tree $j$. The aggregate rate from all services over link $l$ is upper bounded by the available link transmission rate $B(l)$, i.e.,
\begin{equation}
    \sum_{r \in \mathcal{R}}\sum_{j=1}^{J^{r}} \sum_{i =0}^{\left| \mathcal{V}^{r}\right|} \gamma_{li}^{jr} \leq B(l), ~ l \in \mathcal{L}.
    \label{eq05}
\end{equation}

In summary, the optimization problem for the single service scenario is formulated as
\begin{subequations}
    \begin{align}
    & \boldsymbol{\mathrm{(P1):}} \min
     \alpha \sum_{l \in \mathcal{L}}\sum_{j=1}^{J^{r}}\sum_{i=0}^{|\mathcal{V}^{r}|}\!\big(\frac{\gamma_{li}^{jr}}{B(l)} + x_{li}^{jr}\big) + \beta\sum_{i=1}^{|\mathcal{V}^{r}|}\!\sum_{n \in \mathcal{M}}\!\!\frac{C(f_i^{r})}{C(n)}z_{ni}^{r}
     \label{obj}
     \\
    &  \text{subject to}\hspace{0.5cm}
    \eqref{eq01}-\eqref{eq04},
    (\ref{eq11}) - (\ref{eq:subscription_capacity}), (\ref{eq06}), (\ref{eq05})
     \\
    & \hspace{1.9cm}
     \boldsymbol{x},\boldsymbol{y},\boldsymbol{z},\boldsymbol{u},\boldsymbol{\pi} \in \{0,1\},~
    \boldsymbol{d} \succeq 0,~ \boldsymbol{\gamma} \succeq 0.
    \label{eq18c}
    \end{align}
    \label{prob01}%
\end{subequations}
In \eqref{prob01}, the objective function and all constraints are linear except constraints \eqref{eq11}, \eqref{eq06}, \eqref{eq:subscription_p1}. In the next step, we transform these non-linear constraints to equivalent linear constraints such that a standard MILP solver can handle them. To do so, for nonlinear constraint \eqref{eq11}, the bilinear term $\pi^{jr} u_{nit}^{r}$ can be handled by introducing a new variable $w_{nit}^{jr} = \pi^{jr} u_{nit}^{r}$. Constraint \eqref{eq11} is changed to
\begin{align}
    \sum_{(m,n) \in \mathcal{L} }  & y_{(m,n)it}^{jr} - \sum_{(n,m) \in \mathcal{L}} y_{(n,m)it}^{jr} = w_{nit}^{jr} - w_{n(i+1)t}^{jr},
    \nonumber\\
    &\hspace{0.3cm}
    ~n \in \mathcal{N}, i \in \mathcal{S}_0^{|\mathcal{V}^{r}|}, t \in \mathcal{D}^{r}, j \in \mathcal{S}_1^{J^{r}}, r \in \mathcal{R}.
    \label{eq29}
\end{align}
The corresponding relations among $w_{nit}^{jr}$, $\pi^{jr}$, and $u_{nit}^{r}$ are given by
\begin{align}
w_{nit}^{jr} \leq \pi^j,~ & w_{nit}^{jr} \leq u_{nit}^{r},~
w_{nit}^{jr} \geq \pi^{jr} + u_{nit}^{r} - 1,
\nonumber\\
&
~n \in \mathcal{N}, i \in \mathcal{S}_0^{|\mathcal{V}^{r}|}, t \in \mathcal{D}^{r}, j \in \mathcal{S}_1^{J^{r}}, r\in\mathcal{R}.
\label{eq21}
\end{align}
For nonlinear constraint \eqref{eq:subscription_p1}, denote the product term $z_{ni}^{r} k_{ni}$ by $g_{ni}^{r}$. Consequently, \eqref{eq:subscription_p1} can be expressed by
\begin{subequations}
\begin{eqnarray}
    & &\hspace{-1.3cm} g_{ni}^{r} \leq z_{ni}^{r}, ~ n \in \mathcal{M},~i \in \mathcal{S}_{1}^{|\mathcal{V}^{r}|} \label{eq:subscription_p1_linear1} \\
    & &\hspace{-1.3cm} g_{ni}^{r} \leq k_{ni}, ~ n \in \mathcal{M},~i \in \mathcal{S}_{1}^{|\mathcal{V}^{r}|} \label{eq:subscription_p1_linear2} \\
    & &\hspace{-1.3cm} g_{ni}^{r} \geq z_{ni}^{r} + k_{ni} - 1, ~ n \in \mathcal{M},~i \in \mathcal{S}_{1}^{|\mathcal{V}^{r}|}. \label{eq:subscription_p1_linear3}
\end{eqnarray}
\label{eq:subscription_capacity_linearized}%
\end{subequations}
For nonlinear constraint \eqref{eq06}, we utilize the big-$M$ method, and express it equivalently as
\begin{subequations}
    \begin{align}
    \nonumber \label{eq:p2:c12}
    & d^{jr}  -  M(1-x_{li}^{jr}) \leq \gamma_{li}^{jr} \leq d^{jr} , \\
    &
    ~ l \in \mathcal{L}, i \in \mathcal{S}_0^{\left| \mathcal{V}^{r} \right|},~ j \in \mathcal{S}_{1}^{J^{r}},~r \in \mathcal{R}
    \\
    \label{eq:p2:c13}
    & \nonumber 0 \leq\gamma_{li}^{jr}  \leq M x_{li}^{jr}, \\
    & ~ l \in \mathcal{L}, i \in \mathcal{S}_0^{\left| \mathcal{V}^{r} \right|},~ j \in \mathcal{S}_{1}^{J^{r}},~r \in \mathcal{R}
    \end{align}
    \label{eq18}%
\end{subequations}%
\noindent
where $M$ is a large positive number. Since $d^{jr}$ is upper bounded by $\overline{d^{r}}$, $\gamma_{li}^{jr}$ given by \eqref{eq06} is bounded above by $\overline{d^{r}}$. Thus, it suffices to set $M = \overline{d^{r}}$.
%

As a result, the non-linear optimization problem for a single-service in \eqref{prob01} can be re-written in an MILP form as
\begin{subequations}
    \begin{align}
    & \boldsymbol{\mathrm{(P1^{\prime})  }:} \,\, \underset{\boldsymbol{\mathcal{X}}}{\min}
     \sum_{l \in \mathcal{L}}\sum_{j=1}^{J^{r}}\sum_{i=0}^{|\mathcal{V}^{r}|}\!\alpha\bigg(\frac{\gamma_{li}^{jr}}{B(l)} + x_{li}^{jr}\bigg) + \beta\sum_{i=1}^{|\mathcal{V}^{r}|}\!\sum_{n \in \mathcal{M}}\!\!\frac{C(f_i^{r})}{C(n)}z_{ni}^{r}
     \label{eq:MILP:obj}
     \\
    &
    \text{subject to} ~ (\ref{eq01})-(\ref{eq04}),
    (\ref{eq12}) - (\ref{eq:capacity_p1}), 
     (\ref{eq05}),
     (\ref{eq18c})-(\ref{eq18})
    \end{align}\label{eq:MILP}%
\end{subequations}
where $\boldsymbol{\mathcal{X}} = \{\boldsymbol{x}, \boldsymbol{y}, \boldsymbol{z}, \boldsymbol{u}, \boldsymbol{w}, \boldsymbol{\pi}, \boldsymbol{d},\boldsymbol{\gamma}\}$, and \eqref{eq:MILP} can be solved by an MILP solver.



\subsection{Multi-service Multipath Scenario}
\label{sec:problem_formulation:multiservice}
In this subsection, we consider the scenario of jointly handling multiple multicast service requests. We formulate an MILP that jointly constructs the multicast topology for multiple service requests, where the goal is to find the combination of service requests such that the aggregate throughput is maximized while the overall function and link provisioning costs are minimized.
First, we need to maximize the achieved aggregate throughput obtained by hosting network services on the substrate network. The achieved aggregate throughput is given by
\begin{align}
    \mathbb{R} = \sum_{r \in \mathcal{R}} R^{r} \rho^{r},
    \label{eq:R_overall}
\end{align}
where
$\rho^{r} \in \{0,1\}$ is a binary decision variable with $\rho^{r} = 1$ when service $r$ is accepted, and $R^{r}$ is the corresponding throughput, defined as
\begin{equation}
  R^{r} = a_1 \sum_{i=1}^{|\mathcal{V}^{r}|} C(f^{r}_i) + 
         a_2 \Big( |\mathcal{V}^{r}| + |\mathcal{D}^{r}|  \Big) \overline{d^r},~r\in \mathcal{R}
\label{eq:R1}
\end{equation}
where the first and second terms denote the required amount of processing
and transmission rates, respectively, in packet/s \cite{6550275,Ghodsi:2012:MFQ,Ghodsi:2011:DRF}. The two parameters, $a_1$ and $a_2$, are used to tune the priority of processing and transmission rates respectively, where $a_1 + a_2 = 1$ and $a_1,\,a_2 > 0$.

%
%
Second, in addition to maximizing the achieved aggregate throughput, an efficient configuration to utilize the resources efficiently is needed. Specifically, there can be multiple solutions to maximize the aggregate throughput defined above. Among such solutions, we find the configuration with the least function and link provisioning cost to save resources for future services. The multi-service scenario is cast as a two-step MILP, the first step aims to find the maximum achievable aggregate throughput, followed by a formulation which finds the routing and NF placement for each admitted NF chain subject to the maximum achievable aggregate throughput.

In the multi-service scenario, some of the service requests can be rejected due to limited resources. Therefore, we first generalize some of the previous constraints as follows. Constraint \eqref{eq04} is generalized to
\begin{equation}
    \sum_{j=1}^{J^{r}} d^{jr} \geq \rho^{r} \overline{d^{r}}, ~ r \in \mathcal{R}
    \label{eq:p2:c3}
\end{equation}
where the summation of the fractional transmission rate from all trees for service $r$ ($ \in \mathcal{R}$) is forced to zero when the service is rejected (i.e. when $\rho^r=0$). Moreover, an instance of $f_i^r$ is deployed at only one NFV node if service $r$ is accepted, i.e., \eqref{eq08} becomes
\begin{equation}
     \sum_{n \in \mathcal{M}} u_{nit}^{r} = \rho^{r}, ~ i \in \mathcal{S}_1^{\left| \mathcal{V}^{r} \right|}, ~ t \in \mathcal{D}^{r}, ~ r \in \mathcal{R}.
    \label{eq:p2:c6}
\end{equation}
Similarly, when service $r$ is rejected, all variables related to the service should be zero, i.e.,
\begin{equation}
    \pi^{jr} \leq \rho^{r}, ~ z_{ni}^{r} \leq \rho^{r}, ~ n \in \mathcal{N}, i \in \mathcal{S}_1^{\left| \mathcal{V}^{r} \right|},~j \in \mathcal{S}_{1}^{J^{r}},~r \in \mathcal{R}.
    \label{eq:p2:c8}
\end{equation}

\emph{Objectives} --- The objective of the first step is to find the maximum aggregate throughput $\mathbb{R}^{\ast}$. Then, we aim to minimize the function and link provisioning costs for all admitted services, subject to the maximum achievable aggregate throughput $\mathbb{R}^{\ast}$, in the second step.

%
Now we present the first step of maximizing the aggregate throughput as
\begin{subequations}
    \begin{align}
    & \boldsymbol{\mathrm{(P2)}:}   \underset{\boldsymbol{x},\boldsymbol{y},\boldsymbol{z},\boldsymbol{u},\boldsymbol{\rho},\boldsymbol{\pi},\boldsymbol{w},
    \boldsymbol{d},\boldsymbol{r}}{\max} \sum_{r \in \mathcal{R}}R^{r} \rho^{r}
    \\
    \nonumber & \text{subject to}   \\ 
    &  \quad \quad \eqref{eq01}, \eqref{eq03}, \eqref{eq12}, \eqref{eq:capacity_p1}, \eqref{eq05}, \eqref{eq29}, \eqref{eq21} \\
    &  \quad \quad  \eqref{eq18}, \eqref{eq:subscription_capacity_linearized}, \eqref{eq:p2:c3}-\eqref{eq:p2:c8}
    \\
    &  \quad \quad \boldsymbol{x},\boldsymbol{y},\boldsymbol{z},\boldsymbol{u},
    \boldsymbol{\rho},\boldsymbol{\pi},\boldsymbol{w} \in \{0,1\},~ \boldsymbol{d},\boldsymbol{r} \geq 0.
    \end{align}
    \label{prob03}
\end{subequations}%
After solving \eqref{prob03}, we obtain a configuration that provide a maximal aggregate throughput, $\mathbb{R}^{\ast}$, for the given $|\mathcal{R}|$ services and substrate network. However, such configuration can be one among many that can yield $\mathbb{R}^{\ast}$. Among all possible configurations, we choose one such that the function and link provisioning costs are minimized.

%
%
%
Therefore, in the second step, we find the combination of admitted services and their multicast topologies such that the function and link provisioning costs are minimized, subject to the maximum achievable aggregate throughput $\mathbb{R}^{\ast}$, as follows,
\begin{subequations}
    \begin{align}
    &
    \nonumber \boldsymbol{\mathrm{(P3)}:} \,    \underset{\boldsymbol{x},\boldsymbol{y},\boldsymbol{z},\boldsymbol{u},\boldsymbol{\rho},\boldsymbol{\pi},\boldsymbol{w},
        \boldsymbol{d},\boldsymbol{r}}{\min}  \\
        & \sum_{r \in \mathcal{R}}\sum_{l \in \mathcal{L}}\sum_{j \in \mathcal{S}_1^{J ^{r}}}\sum_{i = 0}^{|\mathcal{V}^{r}|}
    \alpha\bigg(\frac{\gamma_{li}^{jr}}{B(l)} + x_{li}^{jr}\bigg)
    + \sum_{r \in \mathcal{R}}\sum_{i=1}^{|\mathcal{V}^{r}|}\sum_{n \in \mathcal{N}}\beta \frac{C(f_i^{r})}{C(n)} z_{ni}^{r}
    \\
    \nonumber & \text{subject to} \\
    & \eqref{eq01}, \eqref{eq03}, \eqref{eq12}, \eqref{eq:capacity_p1}, \eqref{eq:subscription_capacity_linearized}, \eqref{eq05}, \eqref{eq29}, \eqref{eq21}, \eqref{eq18}, \eqref{eq:p2:c3}-\eqref{eq:p2:c8}
    \\
    &  \boldsymbol{x},\boldsymbol{y},\boldsymbol{z},\boldsymbol{u},
    \boldsymbol{\rho},\boldsymbol{\pi},\boldsymbol{w} \in \{0,1\},~ \boldsymbol{d},\boldsymbol{r} \geq 0
    \\
    & \sum_{r \in \mathcal{R}}R^{r}\rho^{r} \geq \mathbb{R}^{\ast}.
    \label{eq102}
    \end{align}
    \label{prob04}%
\end{subequations}
The problem in \eqref{prob04} is an MILP, and can be solved optimally by a standard MILP solver.
After solving \eqref{prob04}, we obtain optimal solutions such that the maximal aggregate throughput $\mathbb{R}^{\ast}$ is achieved with minimal function and link provisioning costs.


\begin{remark}
The joint multicast routing and NF placement problems for both single-service and multi-service cases are NP-hard.
\end{remark}
\emph{Proof:}
We first show that our single-service problem (P1) can be reduced from the Steiner tree problem in polynomial time. Assume a service request with only one function ($f$) and multiple destinations ($\mathcal{D}$). We have a physical substrate ($\mathcal{G}$) such that the source ($s$) is a leaf vertex (in $\mathcal{G}$) connected to the only feasible NFV node for $f$. The optimal solution can be obtained in two steps. First, we build a Steiner tree from the NFV node to the destinations. Second, we place $f$ on the NFV node, and connect the NFV node to the source (as this is the only feasible option). The first step is NP-hard, while the second step is performed in polynomial time. Thus, (P1) is NP-hard. It is then proved that the multi-service problem (P3) is NP-hard as it includes (P1) as a special case \cite{R.Garey1979,Chopra1994}.

\section{Heuristic Algorithms}
\label{sec:heuristic}
Even though $(\mathrm{P1^{\prime}})$, $(\mathrm{P2})$, and $(\mathrm{P3})$ in Section \ref{sec:problem_formulation} can be solved optimally by an MILP solver, the computational time is high. A low-complexity heuristic algorithm is needed to efficiently find a solution. The proposed framework is modular in its design, and can be divided into two main steps. First, a mechanism is employed to prioritize service requests based on some heuristics that aim to maximize the aggregate throughput. Second, the prioritized service requests are embedded sequentially using the joint placement and routing (JPR) algorithm for a single-service.

\subsection{Joint Placement and Routing for Single-service Scenario}
\label{sec:heuristic:single_single}
%
We design a single-service heuristic algorithm based on the following considerations: (i) Different types of NFs can run simultaneously on an NFV node; (ii) The traversed NF types and their order should be considered for each source-destination (S-D) pair; (iii) The objective is to minimize the provisioning cost of the multicast topology based on \eqref{eq14}.
According to the aforementioned principles, a two-step heuristic algorithm is devised as follows:
We first minimize the link provisioning cost by constructing a key-node preferred MST (KPMST)-based multicast topology that connects the source with the destinations; Then, we greedily perform NF placement such that the number of NF instances is minimized.  The pseudo-code of the JPR heuristic is shown in Algorithm \ref{alg:JPR1}, which is explained in more detail as follows.

First, we copy the substrate network $\mathcal{G}$ into $\mathcal{G}'$. Second, to prioritize the NFV node selection in building the KPMST, we calculate new link weights for $\mathcal{G}'$ as
\begin{equation}
\omega_{l'} = \alpha\left(\frac{ \overline{d^r}  }{B(l')}+1\right) + \beta \frac{\overline{d^r}}{C(m)},~l'=(n,m)\in\mathcal{L'}, \mathcal{L'} \subseteq \mathcal{G}',r\in\mathcal{R}
\label{eq:heuCostLink}
\end{equation}
where $C(m)$ is set to a small value when $m$ is a switch; otherwise, it represents the processing resource of NFV node $m$.
Then, a key-NFV node is selected iteratively. We construct the metric closure of $\mathcal{G}'$ as $\mathcal{G}''$, where the metric closure is a complete weighted graph with the same node set $\mathcal{N}$ and with the new link weight over any pair of nodes given by the respective shortest path distance. From the metric closure, we find the MST which connects the source node, destination nodes, and the key-NFV node. An initial multicast routing topology ($\mathcal{G}_v$) can be constructed by replacing the edges in the MST with corresponding paths in $\mathcal{G}'$ wherever needed. We then greedily place the NFs from the source of the multicast topology to its destinations with the objective of minimizing the number of NF instances. The cost $\mathbb{C}(\mathcal{G}_v)$ of the new multicast topology (as in \eqref{eq14}) and the number $\mathbb{A}(\mathcal{G}_v)$ of successfully embedded NF instances are computed. The objective is to jointly maximize the number of successfully placed NFs and minimize the provisioning cost by iterating over all candidate key-NFV nodes. In every iteration, a new key-NFV node is selected. If $\mathbb{A}(\mathcal{G}_v)$ is increased, we update the selected multicast topology with the new key-NFV node; If $\mathbb{A}(\mathcal{G}_v)$ is unchanged and $\mathbb{C}(\mathcal{G}_v)$ is reduced, we also update the selected multicast topology.
If a path cannot accommodate all required NFs ($i.e.$, $f_1,f_2,\dots,f_{|\mathcal{V}|}$) after selecting a key-NFV node, we devise a corrective subroutine that places the missing NF instances on the closest NFV node from the multicast topology, and the corresponding physical links are rerouted as follows. Let $P_{t}$ be the path from $s$ to $t$ in $\mathcal{G}_v$, $P_{t,f}$ be the union of the paths  such that a missing function ($f$) is not hosted \big(i.e.,$ \{ P_{t,f} = \cup_{t\in\mathcal{D}} P_t |\, f \, \mathrm{is\,not\,hosted} \}$\big), and $P_{c,t,f}$ be longest common path before first branch in $P$. Correspondingly, for each missing NF, we link the nearest applicable NFV node to $P_{c,t,f}$, place the missing NF instance, and remove all unnecessary edges.

So far, a flexible multicast topology that connects the source with the destinations, with all NF instances traversed in order, is constructed. We first resume from the single-path scenario ($J=1$) to check whether each path in $G_v$ satisfies the link transmission rate requirement as per \eqref{eq05}, and find an alternative path for each infeasible path. If a single-path solution is infeasible due to \eqref{eq05},
the heuristic algorithm for the single-path case is extended to the multipath-enabled NF chain embedding problem (with $J > 1$). Enabling multipath routing provides several advantages. It is activated when the transmission rate requirement between two consecutively embedded NF instances cannot be satisfied (i.e., when $B(l) < \overline{d^r} ,\,l \in \mathcal{L}$); and multipath routing is to reduce the overall link provisioning cost compared with the single-path case. For each path between two embedded NF instances that does not satisfy the transmission rate requirement $\overline{d^r}$, we increment the number of multipath routes gradually, and look for a feasible solution up to the predefined $J$. The algorithm declares that the problem instance is infeasible when the number of multipath routes exceeds $J$.
%
%
%
%
%
%
For $J>1$, we trigger a path splitting mechanism for each path between two embedded NF instances for each S-D pair as follows.

%
%
Let $W_{i,i+1}^{t,k}$ be the $k$th path between two embedded NFs ($f_i$,$f_{i+1}$) along the network substrate for destination $t$ $(\in \mathcal{D})$, where the cardinality of all such possible paths is $K_{i,i+1}^{t}$. We first rank all candidate paths in a descending order based on the amount of residual transmission resource. Then, we sequentially choose the paths from the candidate paths, such that the summation of all chosen paths' residual transmission rate meets the required transmission rate $\overline{d^r}$.
%
%
Assuming the current number of trees is $j$, the transmission rates allocated on the $k$th path $(W_{i,i+1}^{t,k})$ is then calculated as
\begin{align}
\nonumber \mathbb{R}(W_{i,i+1}^{t,k})= &  \frac{{B_{\min}^k} {\overline{d^r}}} {\sum_{k=1}^{\min\{j,K_{i,i+1}^{t}\}} B_{\min}^k}, \\
 & t\in \mathcal{D},i \in \mathcal{S}_0^{\left| \mathcal{V}\right|}, k \in \mathcal{S}_{1}^{\min\{j,K_{i,i+1}^{t}\}},~r\in\mathcal{R}
\label{multiratio}
\end{align} 
\noindent
where $B_{\min}^k$ is the amount of minimum residual transmission resources for path $W_{i,i+1}^{t,k}$, i.e., $B_{\min}^k = \min_{l \in W_{i,i+1}^{t,k}} B(l)$. Here, the multipath extension method essentially computes a link-disjoint multipath configuration from a single-path route. Therefore, the proposed multipath extension is necessarily prone to the so-called path diminuition problem, in which not all end-to-end multipath-enabled configurations can be devised from a single-path discovery \cite{Abbas:2006:MPD:1356199.1356203,Abbas:2010:PDN:1664760.1664762}.

Future research is needed to incorporate the end-to-end (E2E) delay requirement to achieve quality of service satisfaction. This is a challenging issue since directly expressing the E2E delay as a function of the decision variables of an optimization problem in a closed form is cumbersome. However, given an embedded NF chain, one can model (or measure) the E2E delay, upon which an (iterative) algorithm can be developed to re-adjust the embedding solution of violated NF chains. For instance, in our previous work \cite{8408468}, we propose an analytical E2E packet delay modeling framework, based on queueing network modeling, for multiple embedded NF chains while taking into account the computing and transmission resource sharing. By incorporating the proposed E2E delay model, we will investigate how to develop a delay-aware NF chain embedding algorithm in our future work.

\begin{algorithm}[htbp]
\SetKwInOut{Input}{Input}
\SetKwInOut{Output}{Output}
\underline{Procedure} JPR $(\mathcal{G},{S}^{r})$\;
\Input{$\mathcal{G'}(\mathcal{N'},\mathcal{L'})$, $S^{r}=(s^{r},\mathcal{D}^{r},f_{1}^{r},f_{2}^{r},\dots, f_{|\mathcal{V}|}^{r},\overline{d^r})$}
\Output{$\mathcal{G}_v$}
  $\mathbb{C}_{r} \leftarrow \infty $; $\mathbb{A}_{r} \leftarrow 0$\; 


\For{$n \in \mathcal{M}$}
{
$\mathcal{G}'' \leftarrow \mathrm{MetricClosure}(\mathcal{G}',\{n,s,\mathcal{D}\})$;
$\mathcal{G}_v^{temp} (\mathcal{N}_v,\mathcal{L}_v) \leftarrow \mathrm{KruskalMST}(\mathcal{G}'')$\;


    \lFor{path from $s$ to each $t \in \mathcal{D}$}{
    place NFs from $\mathcal{V}$ on available NFV nodes sequentially subject to \eqref{eq:subscription_capacity}}

  \lIf{$\mathbb{A}(\mathcal{G}_v^{t}) = \mathbb{A}_{r} \, \& \, \mathbb{C}(\mathcal{G}_v^{t}) < \mathbb{C}_{r}$}
    {$\mathcal{G}_v \leftarrow \mathcal{G}_v^{t}$; $\mathbb{C}_{r} \leftarrow  \mathbb{C}(\mathcal{G}_v^{t})$}
  \lElseIf{$\mathbb{A}(\mathcal{G}_v^{t}) > \mathbb{A}_{ref}$ }
  { 
        $\mathcal{G}_v \leftarrow \mathcal{G}_v^{t}$;  $\mathbb{A}_{ref} \leftarrow \mathbb{A}(\mathcal{G}_v^{t})$; $\mathbb{C}_{ref} \leftarrow \mathbb{C}(\mathcal{G}_v^{t})$}}

\lFor{each missing NF ($f$) from $\mathcal{G}_v$}{link nearest NFV node that can host $f$ to $P_{c,t,f}$, and remove unnecessary edges}




    \For{path from $f_{i}$ to $f_{i+1}$ for each $t$ in $\mathcal{G}_{v}$}{
  success $\leftarrow$ false\;
  \For{$j=1:J$}{
    Find $temp = \min\{K_{i,i+1}^{t},j\}$ paths from $\mathcal{G}$\;
    \lIf{$\sum_{k=1}^{temp} \min_{l \in W_{i,i+1}^{t,k}} B(l) \geq \overline{d^r}$}{
    allocate transmission resource for each $k$th path ($W_{i,i+1}^{t,k}$) using \eqref{multiratio}; success $\leftarrow$ true; break}
    
    }
    \lIf{success $=$ false}{break}
    }

\lIf{success = false}{$\mathcal{G}_{v} \leftarrow \mathrm{none}$}
\lElse{return $\mathcal{G}_{v}$}


\caption{Heuristic algorithm for the joint NF placement and routing}
\label{alg:JPR1}
\end{algorithm}

\subsection{Multi-service Scenario}
\label{sec:heuristic:multiservice}
To achieve high throughput and to efficiently utilize the network resources, our key strategy is to selectively prioritize the network services contributing significantly to the aggregate throughput with least provisioning cost. Here, we consider a static algorithm, i.e., service requests are available \textit{a priori}. 
Three principles serving as criteria to prioritize each service for embedding are identified.
%
The first principle is to rank the given network services based on the aggregate throughput, which is defined in \eqref{eq:R1}.
%
%
%
%
%
A network service with higher achieved throughput has higher priority to be embedded in the substrate network, since such service contributes more to the achieved aggregate throughput than a lower ranked service. However, ranking a service based on the throughput alone does not take into account the impact of the provisioning cost. 
%
It is impossible to obtain the exact provisioning cost of embedding a service request prior to the routing and placement process, as the problem itself is NP-hard. However, the relative positions of source and destinations can provide hints on the cost necessitated to host such service. 
Given a network service with the destinations far from each other (i.e., highly distributive), the provisioning cost is large since more physical links and multicast replication points are expected to connect the destinations. Moreover, the distance from the source to destinations is proportional to network service's provisioning cost as a long routing path with a relatively large number of NF instances is needed to establish the multicast topology.
Combining both effects of the distances between destinations and the distance from source to destinations, we define a \textit{distribution level}, denoted by $g^{r}$, as the product of two components. The first component is $A^{r}/A$, where $A$ is the area of the smallest convex polygon that spans all nodes in the network, and $A^{r}$ is the area of the smallest convex polygon that spans all destinations of service $r$.
The ratio $A^{r}/A$ provides an estimate of how dense a set of destinations of one service is in a given area of the network. Note that existing algorithms to determine the convex hull of a set of points and to calculate the area of an arbitrary shape are available \cite{convex_hull}. 
The second component is $q^{r}/q$, where $q^{r}$ is the distance from source to the center point of the set of destinations in service $r$, and $q$ is the largest distance between two arbitrary nodes in the substrate network. The center point of the set of destinations in one service plays a role as a representer for all destinations in that service. 
The ratio $q^{r}/q$ can measure how far is the source from the destinations. The distribution level metric $g^{r}$ is thus expressed as
\begin{equation}
g^{r} = \frac{A^{r}q^{r}}{Aq}, \quad r \in \mathcal{R}.
\label{eq:distribution}
\end{equation}
%
A larger value of $g^{r}$ implies a higher distribution level, where the source is positioned farther away from its destinations and the destinations are more distribution in the whole network coverage area.
%
A largely distributive network service consumes more network resources, resulting in a high provisioning cost. Therefore, a service with a lower value of $g^{r}$ has a higher priority to be embedded in order to preserve the substrate network resources. 
Although the parameters in \eqref{eq:distribution} can be calculated with regard to the hop-count (e.g., via computing the shortest path) which is a more representative metric than the Euclidean distance, it is exhaustive to do so.


Next, we introduce a third metric named \textit{size} to incorporate both \eqref{eq:R1} and \eqref{eq:distribution} as follows,
\begin{equation}
\text{}U^{r}= R^{r} (1-g^r), \quad r \in \mathcal{R}
\label{eq:cost1-1-1-2}
\end{equation}
where the goal is to prioritize a service with higher throughput subject to a correction factor of $1-g^r$ for how distributive (costly) it is.
%

To summarize, we calculate the throughput, the distribution level, and the size for each service request using \eqref{eq:R1}, \eqref{eq:distribution}, and \eqref{eq:cost1-1-1-2}. Then, service requests are sorted according to their sizes in a descending order, and embedded using Algorithm~\ref{alg:JPR1}. 


\subsection{Complexity Analysis}
First, the heuristic algorithm for the single-service scenario (Algorithm \ref{alg:JPR1}) iterates over $|\mathcal{M}|$ NFV nodes to find a key-NFV node. For each potential key-NFV node, a multicast topology is constructed and compared with the previous iteration (in Lines 3-8). Denote the set of destinations, source node, and the key NFV node by $\mathcal{T}$ (i.e., $\mathcal{T} = \{n,s,\mathcal{D}\}$). For each potential key-NFV node, we construct the metric closure on $\mathcal{T}$, which is computed by considering the all-pairs shortest-path algorithm on $\mathcal{T}$. Thus, the worst-case running time is $\mathcal{O}(|\mathcal{N}| |\mathcal{T}|^2)$. Then, we find an MST on the constructed metric closure. The MST is transformed to the Steiner tree by replacing each edge with the shortest path, and removing unnecessary edges (Line 4). The worst-case time complexity of the MST-based Steiner tree is dominated by the metric closure. The construction of the Steiner tree is followed by an NF placement process that requires $\mathcal{O}\{|\mathcal{D}| |\mathcal{M}| \}$ time in the worst-case since a path from the source to each destination passes by at most $|\mathcal{M}|$ NFV nodes (Line 5).
To extend to the multipath scenario, in Lines 10-17, for each path between two embedded NF instances, we find up to $J$ paths and split the traffic according to \eqref{multiratio}, which requires $\mathcal{O}(J \left|\mathcal{D}\right| \left|\mathcal{V}\right| |\mathcal{N}| \log |\mathcal{N}| )$ time in the worst-case. 
For the multi-service scenario, we measure the size of each request, followed by a sorting algorithm based on the size in \eqref{eq:cost1-1-1-2}, which requires $\mathcal{O}( \left|\mathcal{R}\right| \log \left|\mathcal{R}\right| )$ time. Consequently, each service request is embedded sequentially. Hence, the overall worst-case time complexity of the multi-service multi-path scenario is $\mathcal{O}\big(\left|\mathcal{R}\right| \log \left|\mathcal{R}\right| + |\mathcal{N}| |\mathcal{D}| (|\mathcal{D}| +  J \left|\mathcal{V}\right| \log |\mathcal{N}| ) \big)$.

\section{Simulation Results}
\label{sec:simulations}
\begin{figure}[!ht] 
\centering
{\includegraphics[scale=0.35]{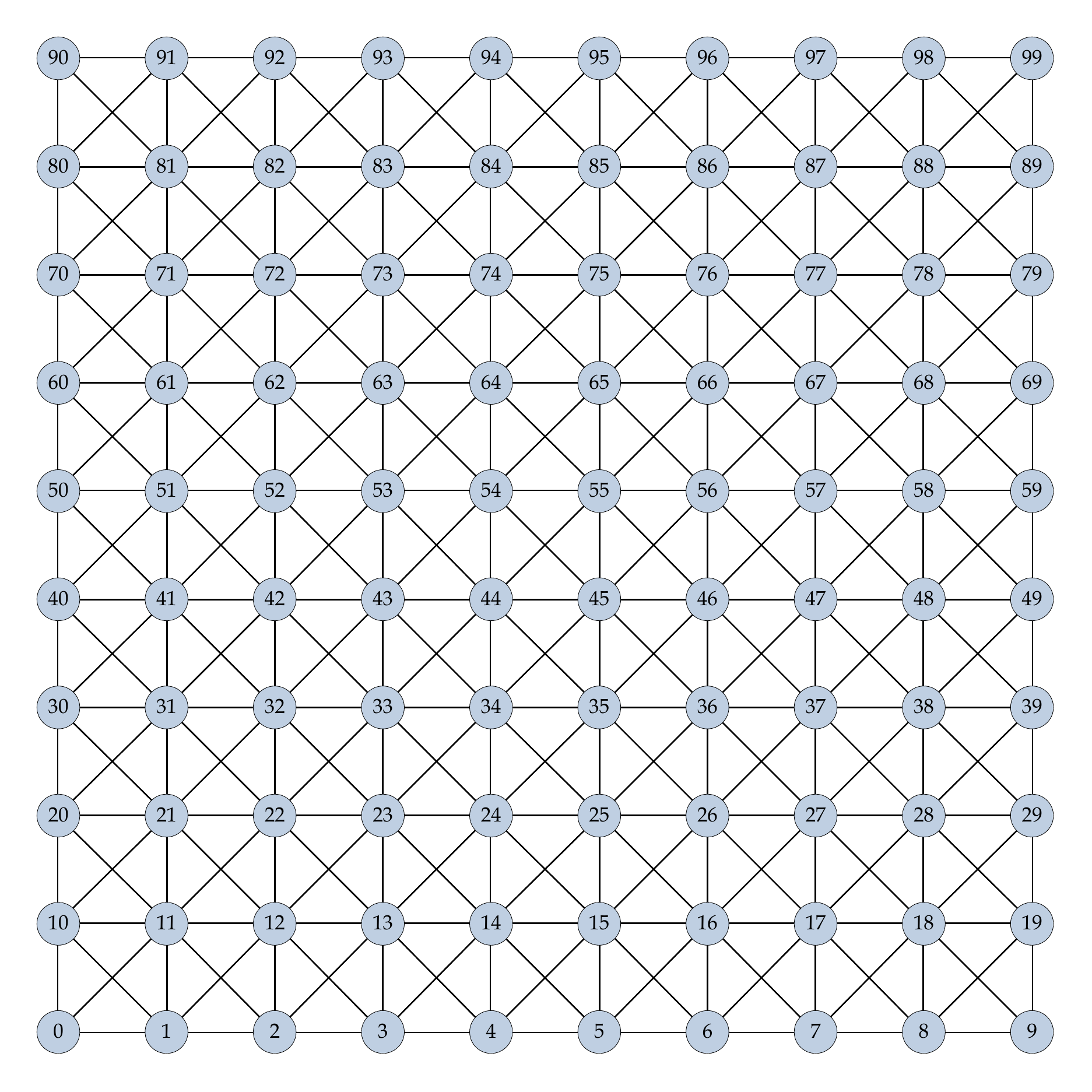}}
\caption{Mesh topology with $|\mathcal{N}|=100$ and $|\mathcal{L}|=684$.}\label{fig:topologies_mesh}
\end{figure}

In this section, simulation results are presented to evaluate the optimal and heuristic solutions to the joint multicast routing and NF placement problems for the single- and multi-service scenarios, considering both single-path and multipath routing cases. The simulated substrate network is a mesh-topology based network \cite{8058426}, with $|\mathcal{N}|=100$ and $|\mathcal{L}|=684$, as shown in Fig. \ref{fig:topologies_mesh}. We randomly choose $25$ vertices as NFV nodes in the mesh network, and the transmission rate of physical link $l$ and the processing rate of NFV node $n$ are uniformly distributed between $0.5$ and $2$ Million packet/s (Mpacket/s), i.e., $B(l), C(n)\sim\mathcal{U}(0.5,2)$ Mpacket/s. To solve the formulated MILP problems, we use the Gurobi solver with the branch and bound mechanism, where the weighting coefficients are set as $\alpha=0.6$, $\beta=0.4$. The processing rate requirement of the NFs are set to be linearly proportional to the incoming data rate, i.e., $C(f^{r}) = \overline{d^r}$ \cite{7417401}.

\begin{figure}[htb]
    \centering
    \begin{subfigure}
        \centering
        \includegraphics[width = 8cm]{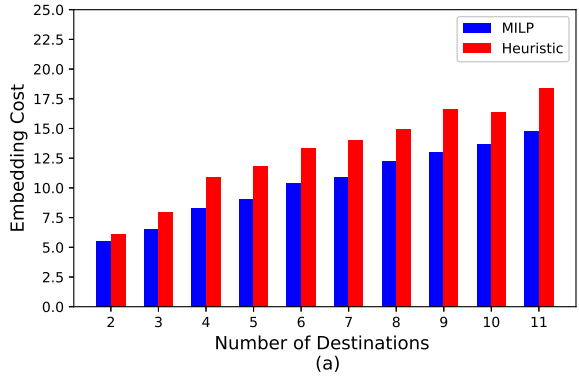}
        \includegraphics[width = 8cm]{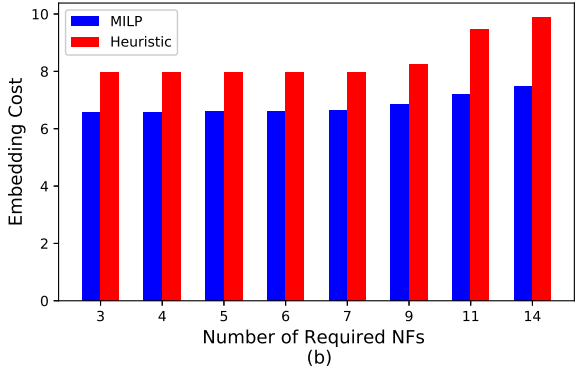}
    \end{subfigure}
    \caption{Embedding cost with respect to (a) the number of destinations and (b) the number of required NFs.}
    \label{fig:single_path_dest_func}
\end{figure}

First, we conduct a comparison between the optimal solution of the single-service single-path MILP formulation and the solution of the heuristic. We generate random service requests where the numbers of NFs and destinations are varied from 3 to 14 and 2 to 11, respectively. The data rate of the generated service requests are set to $\overline{d^r} = 0.2$ Mpacket/s. The total provisioning cost obtained from both optimal and heuristic solutions are shown in Fig. \ref{fig:single_path_dest_func}, as the number of destinations $|\mathcal{D}|$ or NFs $|\mathcal{V}|$ increases. It can be seen that the total provisioning cost increases with $|\mathcal{D}|$ or $|\mathcal{V}|$. As $|\mathcal{D}|$ increases, the costs obtained from both optimal and heuristic solutions grows with nearly a constant gap. Adding destinations incurs a higher cost than adding NF instances, since additional physical links and NF instances are required for each destination. 

\begin{figure}
\centering
\includegraphics[width = 9cm]{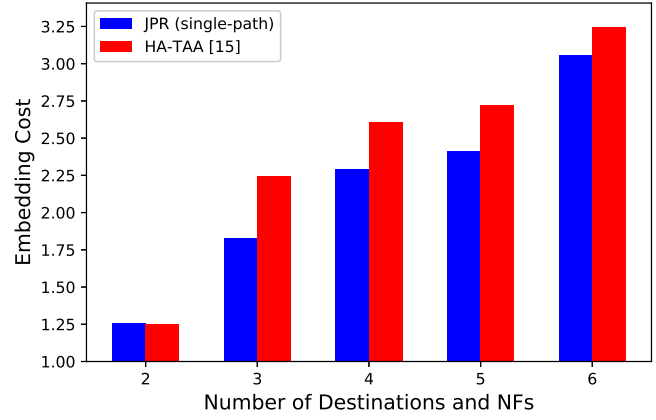}
\caption{Comparison of embedding cost between the proposed JPR heuristic algorithm and HA-TAA heuristic algorithm in \cite{Zhang2016}.}
\label{fig:embedding_cost}
\end{figure}

Next, we compare the proposed heuristic algorithm (JPR) with the heuristic algorithm named HA-TAA\footnote{The heuristic algorithm (HA-TAA) in \cite{Zhang2016} assumes uncapacitated network substrate. We modify the algorithm to the capacitated scenario, change the weights to match our objective function, and assume that an NFV node can host multiple NF instances subject to processing resources. The objective function excludes the minimization of number of hops for fairness.} in \cite{Zhang2016}. For a fair comparison, we consider only the single-path scenario. To add heterogeneity to NFV nodes, among the 25 NFV nodes and up to 6 types of NFs, each NF type can be hosted in an NFV node with the chance of 80\%.
We randomly generate 10 multicast requests that have equal number of NFs and destinations (i.e., $|\mathcal{V}|=|\mathcal{D}|$), and each multicast request is embedded on 30 network substrate instances to reduce the effect of randomness. As shown in Fig. \ref{fig:embedding_cost}, the proposed algorithm consistently outperforms HA-TAA when $|\mathcal{V}|=|\mathcal{D}| > 2$. However, the average running time for the modified HA-TAA algorithm was around 0.115 seconds, whereas our JPR heuristic algorithm took about 2.110 seconds due to the higher time complexity of the involved algorithm. In \cite{Zhang2016}, the HA-TAA algorithm first finds the shortest path from the source to the NFV nodes that can host the required NFs sequentially, and the shortest path from the last NFV node to the respective closest destination. Second, it connects the destinations through an MST. Since the placement of each NF is only dependent on the previous NF, the HA-TAA algorithm may take a long path to place NFs before connecting the last NFV node to the closest destination. 
In our algorithm, we first focus on finding a Steiner tree from the source to the destinations through a key-NFV node, thereby optimizing the link provisioning cost. Then, we modify some of the links to host all required NF instances (if needed). Here, NFs can be duplicated to serve different sets of destinations, thereby providing more flexibility and reduced overall provisioning cost. When $|\mathcal{V}|\leq 2$, the HA-TAA is specifically more efficient as the NF placement is related to the locations of both the source and the closest destination.

\begin{figure}
\centering
\includegraphics[width = 9cm]{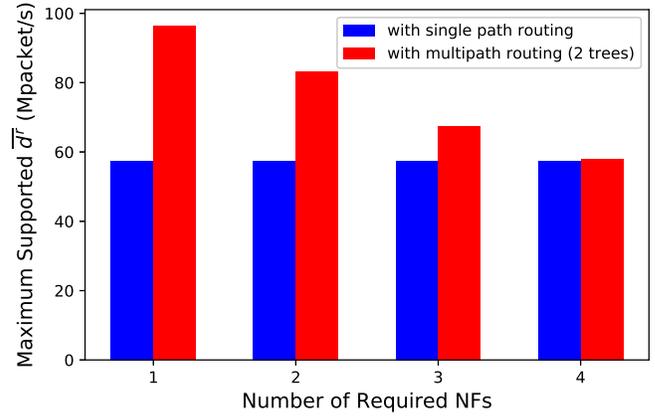}
\caption{Maximum supported data rate ($\overline{d^r}$) for both single-path and multipath routing scenarios using the proposed optimal formulation.}
\label{fig:supported_rate}
\end{figure}

Fig. \ref{fig:supported_rate} shows the advantage of multipath routing over single-path routing using the proposed optimal formulation. We depict the maximum supported data rate $\overline{d^r}$, with which the NF embedding is feasible. 
Compared to the single-path routing case, multipath routing $(J=2)$ always supports higher or equal data rates. With an increase of the number of required NFs, the maximum supported data rate decreases, and converges to the single-path scenario; the processing cost becomes more significant and the number of candidate NFV nodes and paths decrease.

\begin{figure}[htbp]
  \centering
  \includegraphics[width = 7.5cm]{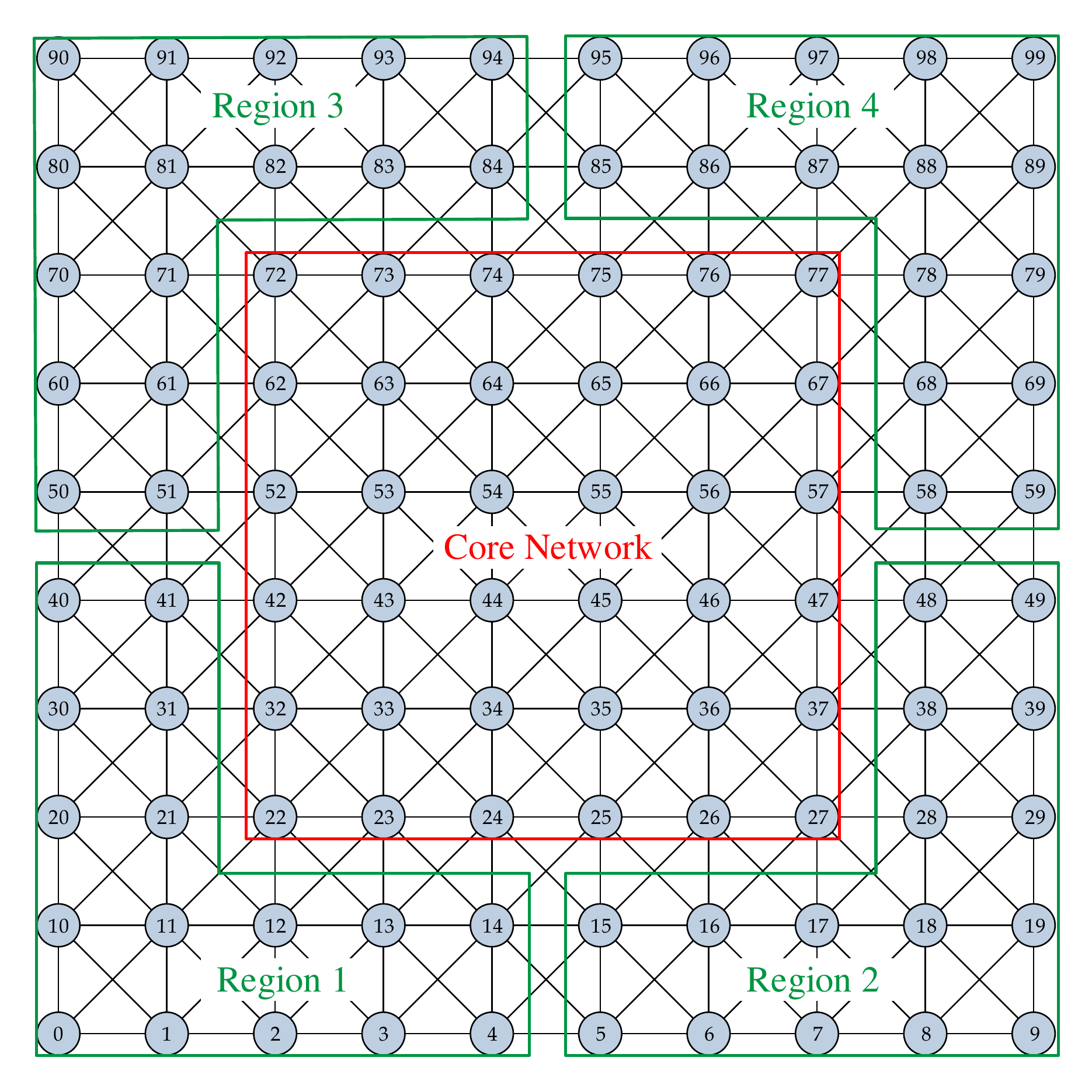} %
  \caption{Mesh topology with $|\mathcal{N}| = 100$, $|\mathcal{L}| = 684$, and 4 access regions and 1 core network region.}
  \label{fig:mesh_topology}
\end{figure}

\begin{table}[htbp]
\centering
\caption{Processing and transmission rates}
\label{Table2}
\begin{tabular}{c|ccc}
\hline\noalign{\vskip 0.3mm}\hline\noalign{\smallskip}
Scenarios & Processing rate & Link transmission & NFV nodes\\
\noalign{\smallskip}\hline\noalign{\smallskip}
Scenario 1 & $\mathcal{U}(3, 8)$ Mpacket/s & $\mathcal{U}(3, 8)$ Mpacket/s  & 47\\
Scenario 2 & $\mathcal{U}(4, 9)$ Mpacket/s & $\mathcal{U}(4, 9)$ Mpacket/s  & 50\\
Scenario 3 & $\mathcal{U}(5, 10)$ Mpacket/s & $\mathcal{U}(5, 10)$ Mpacket/s & 53\\
\hline\noalign{\vskip 0.3mm}\hline
\end{tabular}
\end{table}
Next, we demonstrate the effectiveness of the proposed heuristic admission mechanism for the multi-service scenario discussed in Subsection \ref{sec:heuristic:multiservice}. First, we divide the mesh topology into four access network regions and one core network region as indicated in Fig. (\ref{fig:mesh_topology}). Three scenarios with different available processing and transmission rates on the NFV nodes and physical links are considered in the scale of Mpacket/s, as listed in Table \ref{Table2}.
Each service randomly originates from one access network region, traverses the core network region, and terminates in one of the other three access network regions. For each network scenario, to simulate network congestion, 35 multicast service requests are randomly generated and submitted for embedding, where the data rate $\overline{d^r}$ of service $r$ is randomly distributed between $[1.5, 3.5]$ Mpacket/s, and the number of NFs and destinations are randomly generated as $|\mathcal{V}^r| =\{3,4\}$ and $|\mathcal{D}^r| = \{3,4,5\}$.
For each scenario, the service generation and embedding are repeated 5 times to reduce the effect of randomness. As a benchmark, we use a second strategy that randomly selects the service requests for embedding.

\begin{figure}[htbp]
  \centering
  \includegraphics[width = 8.5cm]{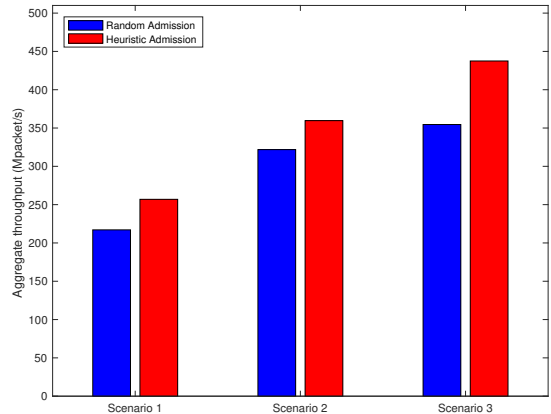} %
  \caption{Aggregate throughput comparison of the random admission and the proposed heuristic admission.}
  \label{heuristicRevenue}
\end{figure}

\begin{figure}[htbp]
  \centering
  \includegraphics[width = 8.5cm]{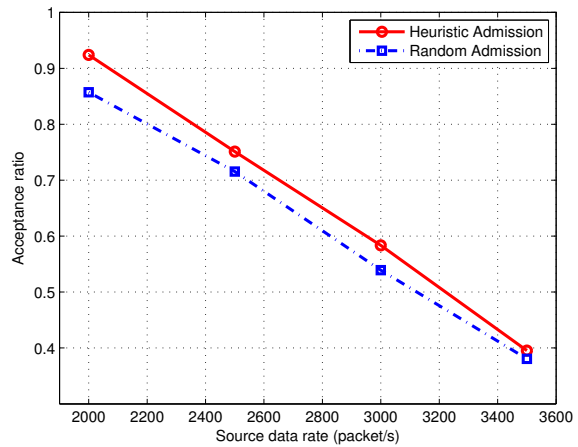}
  \caption{Acceptance ratio comparison of the random admission and the proposed heuristic admission.}
  \label{heuristicRatio}
\end{figure}

\begin{figure}[htbp]
  \centering
  \includegraphics[width = 8.5cm]{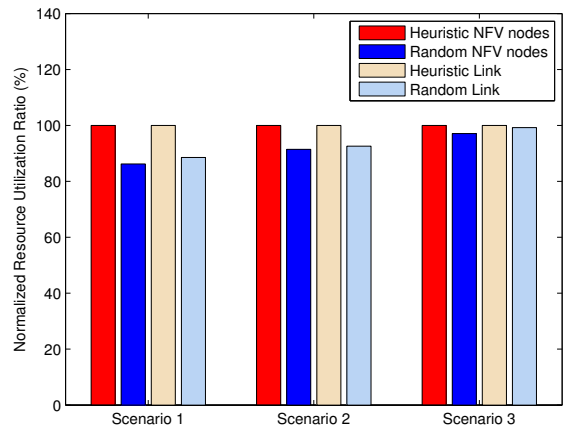} %
  \caption{Resource utilization ratio comparison of the random admission and the proposed heuristic admission.}
  \label{heuristicUti}
\end{figure}
%
%
Fig. \ref{heuristicRevenue} shows the aggregate throughput $\mathbb{R}$ as in \eqref{eq:R_overall} achieved by the random and heuristic admission solutions under three scenarios specified in Table \ref{Table2}, which increases as the available processing and transmission rates increase.
As shown, the aggregate throughput of the proposed heuristic solution exceeds the aggregate throughput of the random admission by 15.63\% on average over all the scenarios. This is because the size metric used in the heuristic solution ensures that the services with a larger throughput are embedded with a higher priority.
Fig. \ref{heuristicRatio} shows the acceptance ratio of the total 35 service requests under different average data rates. The acceptance ratio of the heuristic solution exceeds the random solution over all source data rate levels by 4\% on average.

Fig. \ref{heuristicUti} compares the normalized resource utilization between the random and heuristic admission solutions.
The utilization ratio is calculated as the amount of resources consumed by all embedded services over the total available resources of the substrate network.
We normalize the measured utilization ratio by the utilization ratio of the proposed heuristic admission to highlight the enhancement brought by the heuristic solution. 
In scenarios 1 and 2, the heuristic solution increases the utilization ratio by approximately 12.62\% and 7.97\% over that of the random admission solution, respectively.
In scenario 3, the difference in the utilization ratio between the two solutions decreases, especially the link transmission usage. Recall that the heuristic scheme aims to maximize the aggregate throughput as defined in \eqref{eq:R_overall}. Therefore, although in scenario 3 the utilization ratio achieved by the heuristic scheme is close to that of the random scheme, the aggregate throughput achieved by the former is larger as shown in Fig. \ref{heuristicRevenue}.

\section{Conclusion}
\label{sec:conclusions}
In this paper, we study a joint traffic routing and NF placement framework for multicast services over a substrate network under an SDN-enabled NFV architecture. Within the framework, joint multipath-enabled multicast routing and NF placement is investigated first for a single-service scenario and then for a multi-service scenario. For the single-service scenario, an optimization problem is formulated to minimize function and link provisioning cost, under the physical resource constraints and flow conservation constraints. Our problem formulation is flexible as it allows one-to-many and many-to-one NF mapping, and incorporates multipath routing by constructing multiple trees to deliver the multicast service. The formulated problem is an MILP, and thus can be solved to obtain optimal solutions as benchmark. For the multi-service case, we present an optimization framework that jointly deals with multiple service requests. An optimal combination of service requests and their joint routing and NF placement configurations are studied, such that the aggregate throughput of the core network is maximized, while the function and link provisioning costs are minimized. To reduce the computational complexity in solving the problems in both scenarios, heuristic approaches are proposed to find accurate solutions close to that of the optimal solutions. Simulation results are presented to demonstrate the effectiveness and accuracy of the proposed heuristic algorithms.


%



\section*{Acknowledgments}
This work was supported by research grants from Huawei Technologies Canada and from the Natural Sciences and Engineering Research Council (NSERC) of Canada.

\ifCLASSOPTIONcaptionsoff
  \newpage
\fi



%
\bibliographystyle{myIEEEtran}
\bibliography{references_final}




%

\begin{IEEEbiography}[{\includegraphics[width=1in,height=1.25in,clip,keepaspectratio]{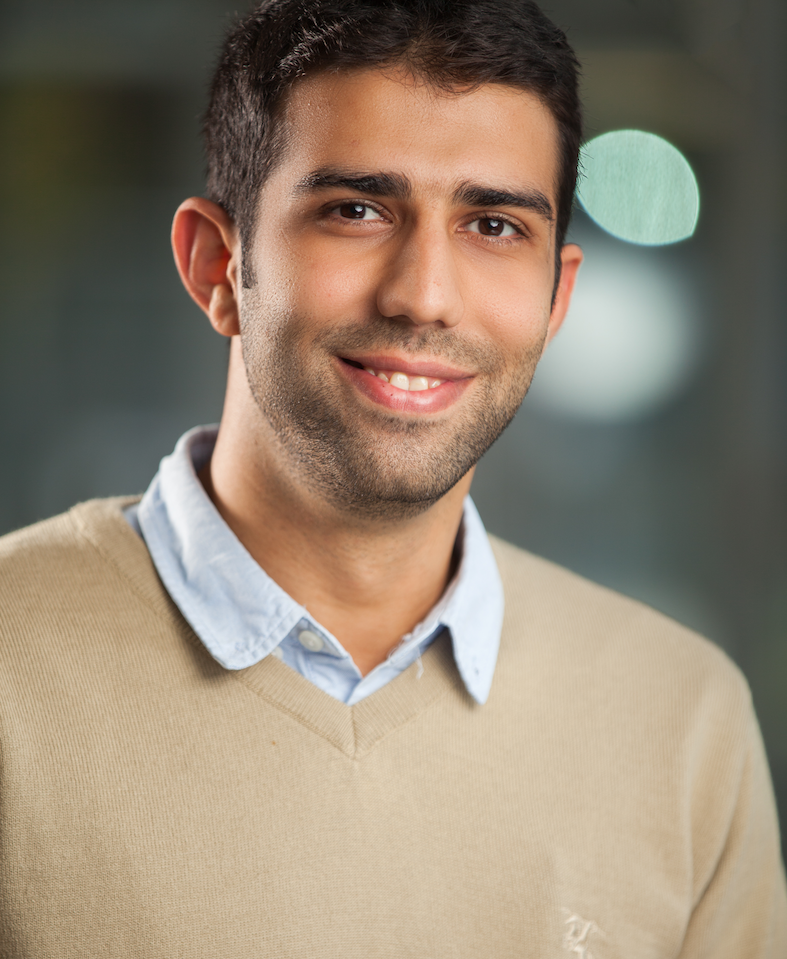}}]{Omar Alhussein}
is currently persuing the Ph.D. degree in electrical engineering with the University of Waterloo, Ontario, Canada. He received the M.A.Sc. degree in engineering science from Simon Fraser University, British Columbia, Canada, in 2015, and the B.Sc. degree in communications engineering from Khalifa University, Abu Dhabi, United Arab Emirates, in 2013. His research interests include next generation wireless networks, network (function) virtualization, wireless communications, and machine learning.
\end{IEEEbiography}

\begin{IEEEbiography}[{\includegraphics[width=1in,height=1.25in,clip,keepaspectratio]{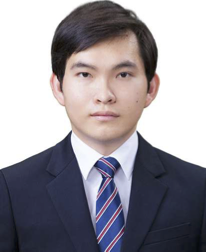}}]{Phu Thinh Do}
received the B.S. degree in electrical engineering from the Ho Chi Minh City University of Technology in 2010 and the M.S.E. and Ph.D. degrees from the Department of Electronics and Radio Engineering, Kyung Hee University, Korea, in 2016. From 2016 to 2019, he was first a postdoctoral fellow at the Digital Communication Lab, Kyung Hee University, then at the Broadband Communications Research Lab, University of Waterloo, Canada. He is now a researcher at the Institute of Research and Development, Duy Tan University, Vietnam. His current research interests include 5G wireless communications, search engines and recommendation systems for e-commerce platforms.
\end{IEEEbiography}

\begin{IEEEbiography}
[{\includegraphics[width=1in,height=1.25in,clip,keepaspectratio]{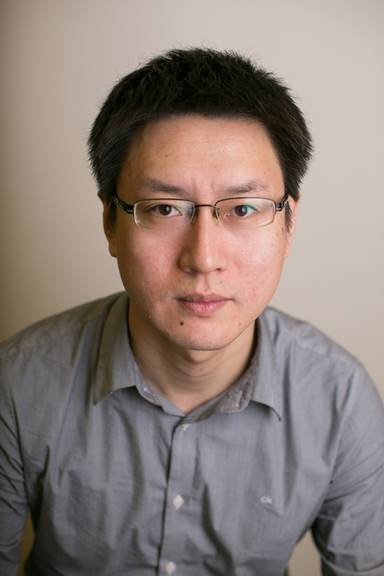}}]
{Qiang Ye} (Member, IEEE) received the Ph.D. degree in electrical and computer engineering from the University of Waterloo, Waterloo, ON, Canada, in 2016. He was a Postdoctoral Fellow with the Department of Electrical and Computer Engineering, University of Waterloo, from December 2016 to November 2018, where he was a Research Associate from December 2018 to September 2019. He has been an Assistant Professor with the Department of Electrical and Computer Engineering and Technology, Minnesota State University, Mankato, MN, USA, since September 2019. His current research interests include 5G networks, software-defined networking and network function virtualization, network slicing, artificial intelligence and machine learning for future networking, protocol design, and end-to-end performance analysis for the Internet of Things.
\end{IEEEbiography}

\begin{IEEEbiography}
[{\includegraphics[width=1in,height=1.25in,clip,keepaspectratio]{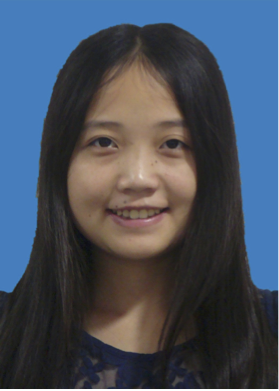}}]
{Junling Li} (IEEE S'18) received the B.S. degree from Tianjin University, Tianjin, China, in 2013, and the M.S. degree from the Beijing University of Posts and Telecommunications, Beijing, China, in 2016. She is currently pursuing the Ph.D. degree with the Department of Electrical and Computer Engineering, University of Waterloo, Waterloo, ON, Canada. Her interests include SDN, NFV, network slicing for 5G networks, machine learning, and vehicular networks. She received the Best Paper Award at the IEEE/CIC International Conference on Communications in China (ICCC) in 2019.
\end{IEEEbiography}

\begin{IEEEbiography}[{\includegraphics[width=1in,height=1.25in,clip,keepaspectratio]{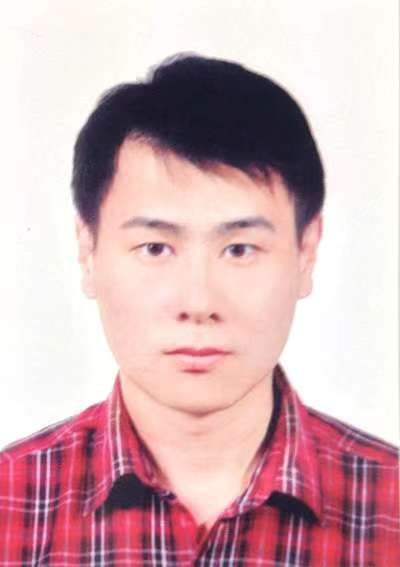}}]{Weisen Shi}
(IEEE S'15) received the B.S. degree from Tianjin University, Tianjin, China, in 2013, and the M.S. degree from the Beijing University of Posts and Telecommunications, Beijing, China, in 2016. He is currently pursuing the Ph.D. degree with the Department of Electrical and Computer Engineering, University of Waterloo, Waterloo, ON, Canada. His interests include drone communication and networking, network function virtualization, and vehicular networks.
\end{IEEEbiography}

\begin{IEEEbiography}[{\includegraphics[width=1in,height=1.25in,clip,keepaspectratio]{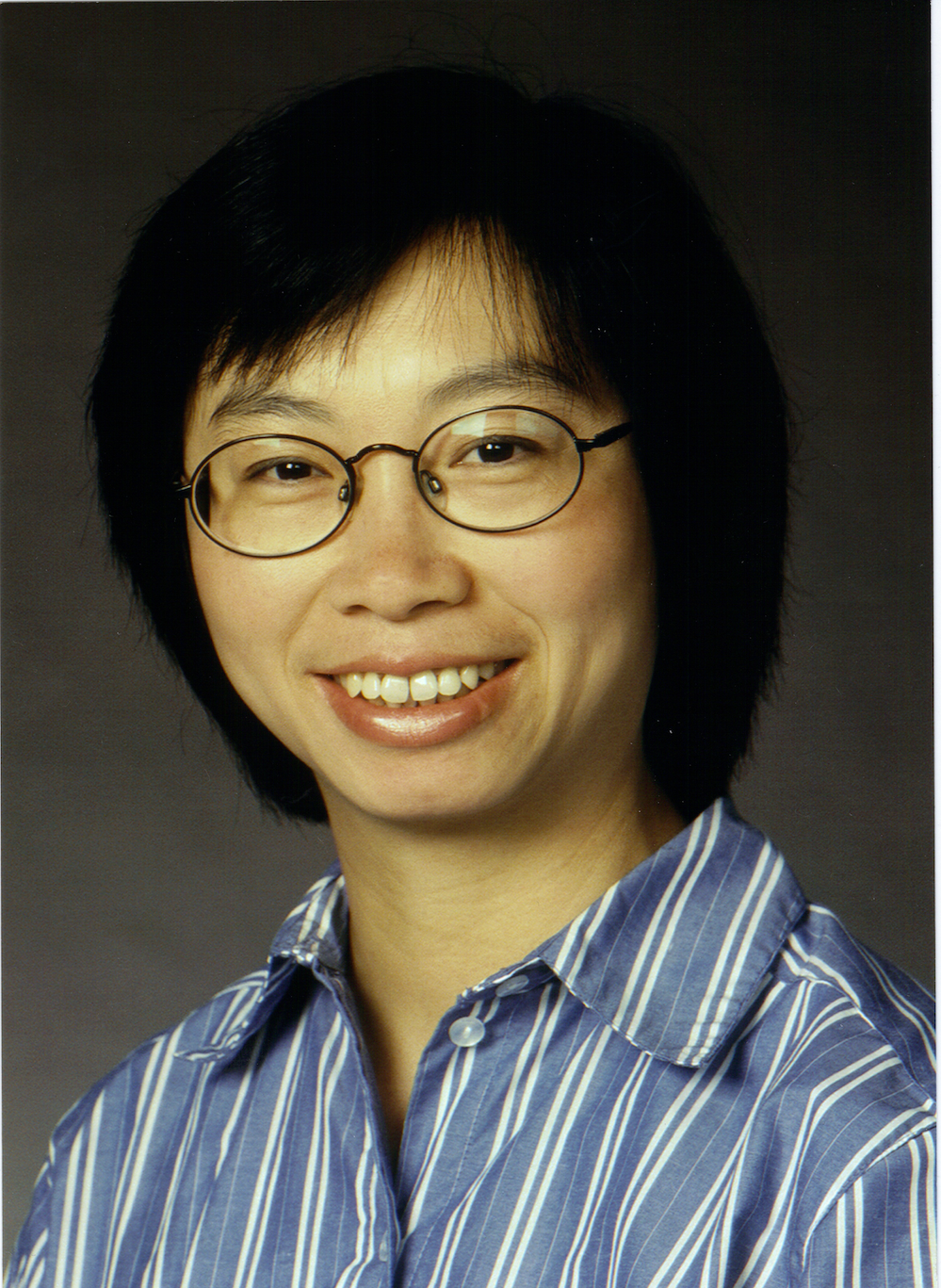}}]{Weihua Zhuang}
(M'93$-$SM'01$-$F'08) has been with the Department of Electrical and Computer Engineering, University of Waterloo, Canada, since 1993, where she is a Professor and a Tier I Canada Research Chair in wireless communication networks. She was a recipient of the 2017 Technical Recognition Award from the IEEE Communications Society Ad Hoc and Sensor Networks Technical Committee, and several best paper awards from IEEE conferences. She was the Technical Program Chair/Co-Chair of the IEEE VTC 2016 Fall and 2017 Fall, the Editor-in-Chief of IEEE Transactions on Vehicular Technology (2007$-$2013), and an IEEE Communications Society Distinguished Lecturer (2008$-$2011). Dr. Zhuang is a Fellow of the Royal Society of Canada, the Canadian Academy of Engineering, and the Engineering Institute of Canada. She is an Elected Member of the Board of Governors and VP Publications of the IEEE Vehicular Technology Society.
\end{IEEEbiography}

\begin{IEEEbiography}[{\includegraphics[width=1in,height=1.25in,clip,keepaspectratio]{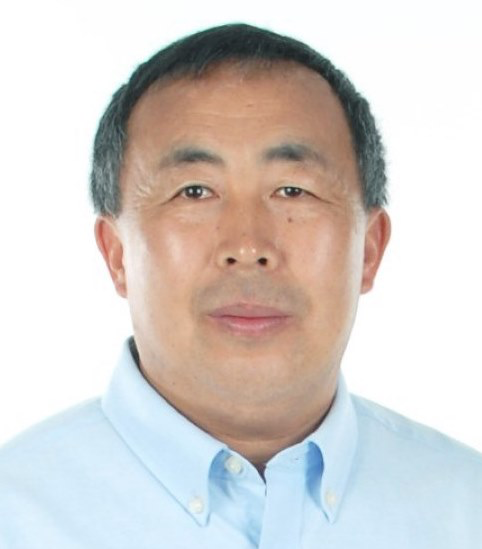}}]{Xuemin (Sherman) Shen}
(M'97-SM'02-F'09) received the Ph.D. degree in electrical engineering from Rutgers University, New Brunswick, NJ, USA, in 1990. He is currently a University Professor with the Department of Electrical and Computer Engineering, University of Waterloo, Canada. His research focuses on network resource management, wireless network security, social networks, 5G and beyond, and vehicular ad hoc and sensor networks. He is a registered Professional Engineer of Ontario, Canada, an Engineering Institute of Canada Fellow, a Canadian Academy of Engineering Fellow, a Royal Society of Canada Fellow, a Chinese Academy of Engineering Foreign Fellow, and a Distinguished Lecturer of the IEEE Vehicular Technology Society and Communications Society.\\ 
Dr. Shen received the R.A. Fessenden Award in 2019 from IEEE, Canada, James Evans Avant Garde Award in 2018 from the IEEE Vehicular Technology Society, Joseph LoCicero Award in 2015 and Education Award in 2017 from the IEEE Communications Society. He has also received the Excellent Graduate Supervision Award in 2006 and Outstanding Performance Award 5 times from the University of Waterloo and the Premier's Research Excellence Award (PREA) in 2003 from the Province of Ontario, Canada. He served as the Technical Program Committee Chair/Co-Chair for the IEEE Globecom'16, the IEEE Infocom'14, the IEEE VTC'10 Fall, the IEEE Globecom'07, the Symposia Chair for the IEEE ICC'10, the Tutorial Chair for the IEEE VTC'11 Spring, and the Chair for the IEEE Communications Society Technical Committee on Wireless Communications. He was the Editor-in-Chief of the IEEE INTERNET OF THINGS JOURNAL and the Vice President on Publications of the IEEE Communications Society.
\end{IEEEbiography}

\begin{IEEEbiography}[{\includegraphics[width=1in,height=1.25in,clip,keepaspectratio]{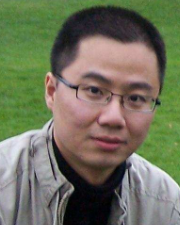}}]{Xu Li}
is a senior principal researcher at Huawei Technologies Canada. He received a PhD (2008) degree from Carleton University, an MSc (2005) degree from the University of Ottawa, and a BSc (1998) degree from Jilin University, China, all in computer science. Prior to joining Huawei, he worked as a research scientist (with tenure) at Inria, France. His current research interests are focused in 5G and beyond. He contributed extensively to the development of 3GPP 5G standards through 90+ standard proposals. He has published 100+ refereed scientific papers and is holding 40+ issued US patents.
\end{IEEEbiography}

\begin{IEEEbiography}
[{\includegraphics[width=1in,height=1.25in,clip,keepaspectratio]{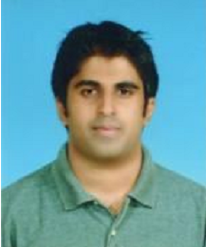}}]
{Jaya  Rao} (jaya.rao@huawei.com) received his B.S. and M.S. degrees in Electrical Engineering from the University at Buffalo, New York, in 2001 and 2004, respectively, and his Ph.D. degree from the University of Calgary, Canada in 2014. He is currently a Senior Research Engineer at Huawei Technologies Canada, Ottawa. Since joining Huawei in 2014, he has worked on research and design of URLLC, V2X and Industrial IoT (IIoT) based solutions in 5G New Radio (NR). He has participated in 5G system impact analysis and performance assessments for contributing towards commercial alliance bodies such as NGMN. He has also contributed at 3GPP RAN WG2, RAN WG3 and SA2 meetings (Rel-15 \& Rel-16) on topics related to URLLC, NR V2X, IIoT, Network Slicing, Mobility Management and Session Management. From 2004 to 2010, he was a Senior Research Engineer at Motorola Inc. He was a recipient of the Best Paper Award at IEEE WCNC 2014.
\end{IEEEbiography}







\end{document}